\documentclass{pasj00}
\draft

\begin{document}
\SetRunningHead{Author(s) in page-head}{Running Head}
\Received{}
\Accepted{}

\title{Growth of a Protostar and a Young Circumstellar Disk with High Mass Accretion Rate onto the Disk}

\author{Takuya \textsc{ohtani} and Toru \textsc{tsuribe} }%
\affil{Department of Earth and Space Science, Graduate School of Science, Osaka University, 1-1 Machikaneyama-cho, Toyonaka, Osaka 560-0043}
\email{ohtani@vega.ess.sci.osaka-u.ac.jp, tsuribe@vega.ess.sci.osaka-u.ac.jp}


%

\maketitle

\begin{abstract}
The growing process of both a young protostar and a circumstellar disk is investigated. 
Viscous evolution of a disk around a single star is considered 
with a model where a disk increases its mass by dynamically accreting envelope 
and simultaneously loses its mass via viscous accretion onto the central star. 
We focus on the circumstellar disk with high mass accretion rate onto the disk $\dot{M}=8.512c_{\rm s}^3/G$ 
as a result of dynamical collapse of rotating molecular cloud core. 
We study the origin of the surface density distribution and the origin of the disk-to-star mass ratio 
by means of numerical calculations of unsteady viscous accretion disk in one-dimensional axisymmetric model. 
It is shown that the radial profiles of the surface density $\Sigma$, azimuthal velocity $v_{\phi}$, 
and mass accretion rate $\dot{M}$ in the inner region approach to the quasi-steady state. 
Profile of the surface density distribution in the quasi-steady state is determined as a result of angular momentum transport rather than its original distribution of angular momentum in the cloud core. 
It is also shown that the disk mass becomes larger than the central star in the long time limit as long as temporary constant mass flux onto the disk is assumed. 
After the mass infall rate onto the disk declines owing to the depletion of the parent cloud core, 
the disk-to-star mass ratio $M_{\rm disk}/M_*$ decreases. 
The disk-to-star mass ratio becomes smaller than unity after $t> 10^5~\rm yr$ and $t>10^6~\rm yr$ from the beginning of the accretion phase in the case with $\alpha_0 =1~{\rm and}~0.1$, 
respectively, where $\alpha_0 $ is the constant part of viscous parameter. 
In the case with $\alpha_0 \leq 10^{-2}$, $M_{\rm disk}/M_*$ is still larger than unity at $2~\rm Myr$ from the beginning of the accretion phase. 
\end{abstract}

\section{Introduction}
It is widely believed that stars are born with circumstellar disks in molecular cloud cores. 
According to the observation, Class I or Class II protostars are typically associated by a disk 
with $10^{-3}-0.1$ ${\rm M}_{\odot}$ (\cite{Bec}). 
In the preceding Class 0 phase, star forming regions are deeply surrounded by an infalling envelope. 
In order to understand the origin of mass in disk and star, an appropriate model for formation and evolution of the disk is desirable. 
Molecular cloud cores are expected to have an angular momentum, 
and cores are thought to collapse to form stars. 
If a cloud core has a sufficient angular momentum, 
infalling gas cannot be directly accreted by the central star because of centrifugal barrier. 
In such a case, disk formation is a natural consequence of the conservation of angular momentum. 
Around a central star, a circumstellar disk and an infalling envelope are expected. 
If one assumes the conservation of angular momentum with a typical amount, 
a small central star and a massive disk are expected as an equilibrium state 
(e.g., \cite{Hayashidisk}). 
However, observations indicate the lower mass disk than above equilibrium state. 
Therefore, some mechanisms are required to decrease the mass of the disk. 
Efficient transport of angular momentum will drive the mass accretion flow onto the central star to decrease mass of the disk. 
This accretion flow is important to determine the disk-to-star mass ratio. 

We assume that a single star and a circumstellar disk forms in an isolated environment. 
In the early phase, only the central part of the molecular cloud core collapses in runaway fashion to form a seed of adiabatic core (runaway collapse phase). 
In the later phase, the envelope falls onto the core and the disk (accretion phase). 
Between above two phases, a seed of circumstellar disk forms and begins to grow. 
The mass accretion rate $\dot{M}$ onto the core and the disk can be written in the form of $\dot{M}=m_0c_{\rm s}^3/G$ with a coefficient $m_0$. 
The coefficient $m_0$ depends on the detail of the collapse. 
For example, $m_0=46.915$ corresponds to an isothermal spherically symmetric self-similar solution by Larson (\yearcite{Larson}). 
On the other hand, $m_0=0.975$ corresponds to another self-similar solution by Shu (\yearcite{shu1977}). 
The solution by Larson (\yearcite{Larson}) (also Penston \yearcite{Penston}) is useful to describe a runaway collapse with dynamical inflow outside. 
On the other hand, 
the solution by Shu (\yearcite{shu1977}) describes the inside-out collapse with static gas sphere outside. 
Rotation of the cloud core is ignored in above two solutions. 
As a solution of collapse with rotation, an axisymmetric self-similar solution for inviscid isothermal flow is studied by \citet{SH98} (hereafter SH98) (see also Narita, Hayashi, \& Miyama \yearcite{NHM84}). 
This solution represents both the runaway collapse of rotating cloud core and the inside-out growth of the disk. 
According to the SH98 solution, in the accretion phase a centrifugally supported disk grows inside the accretion shock which propagates outwards with time. 
Mass accretion rate in SH98 corresponds to $m_0=6$--$10$ at the shock front. 
The precise value of $m_0$ depends on the speed of rotation. 
In the runaway collapse phase, SH98 solution with rotation is similar to Larson's solution with different accretion rate. 
However, in the accretion phase SH98 solution is qualitatively different from spherically symmetric cases. 
With rotation, mass accretion onto the center vanishes since infalling envelope stops at the shock front with finite radius. 
As a result, in the accretion phase, disk with surface density $\Sigma \propto r^{-1}$ forms inside the shock front. 
This surface density distribution is same as equilibrium disks (c.f., \cite{Hayashidisk}; \cite{Mestel}). 

Although above SH98 solution is useful to represent formation and growth of the disk, it is still unrealistic in two points. 
First, since above self-similar solution assumes isothermal equation of state, 
it is not reasonable approximation in the central region where density is sufficiently large and cooling is inefficient. 
In this paper non-isothermal evolution is taken into account. 
Second, in SH98 the effect of angular momentum transport is ignored. 
Turbulent viscosity induced by magneto-rotational instability (MRI) (\cite{BalbusHawley}) is believed to be important 
although it is well known that the molecular viscosity is too small in a circumstellar disk. 
Angular momentum is also possibly transported by the gravitational torque induced by gravitational instability (GI). 
In this paper, as the simplest treatment for angular momentum transport by MRI and GI, 
simple prescription with $\alpha$-viscosity (\cite{Pringle}, \cite{Shakura}) is adopted. 
Results of detailed numerical calculations without viscous prescription are sometimes interpreted using viscous parameters. 
For example, Kratter et al. (\yearcite{Kratter}) indicates that the effective parameter $\alpha$ can be close to unity in gravitationally unstable disk. 
Viscous effect are expected to become important especially in the accretion phase even if it is inefficient during the runaway collapse phase (\cite{NoMine}). 
In the accretion phase, effective viscosity will generate considerable accretion flow from a circumstellar disk to the central star, and it will affect the disk-to-star mass ratio. 
Thus, it is essential to investigate viscous evolution of a disk in the accretion phase. 

To consider the growth of the central star and the surrounding disk as a function of time, unsteady treatment is required. 
Unsteady viscous accretion disk with constant $\alpha$ is studied in the view point of isothermal self-similar solution by Tsuribe (\yearcite{Tsu}) (hereafter T99). 
In T99 solution with the effect of viscosity, different from SH98, 
the central star grows by mass accretion from the disk. 
The radial profile of the surface density in T99 solution is given by $\Sigma\propto r^{-1.5}$ in the inner disk and $\Sigma\propto r^{-1}$ in the outer disk. 
In the inner disk, mass of the disk is small and 
gravitational force from the central star dominates. 
While in the outer disk, considerable disk gravity changes rotation law to the flat rotation ($v_{\phi}= {\rm const}$). 
The effect of non-isothermal temperature distribution considered in this paper modifies the surface density and velocity profiles in the disk. 
 
The profile of the surface density in the disk is important to consider the succeeding planet formation. 
Kokubo \& Ida (\yearcite{KI2002}) (hereafter KI02) discussed that the initial profile of dust surface density 
influences on the timescale of accumulation of planetesimal as well as on the mass of protoplanets. 
They argued that the initial profile of the surface density of protoplanetary disk has the critical power index $\beta =2~{\rm in}~\Sigma\propto r^{-\beta}$. 
In the case with $\beta<2$, mass of protoplanets increase with radius, while it decreases with radius in the case with $\beta>2$. 
However, they do not discuss the origin of the surface density profile of disk. 
Thus, provided that dust to gas ratio is constant, understanding the disk surface density distribution will help understand not only the star formation but also the planet formation. 

Mass accretion rate onto the disk will decline with time owing to the depletion of the parent cloud core. 
Previous self-similar solutions did not include this effect. 
On the other hand, numerical studies by e.g., 
Rice et al. (\yearcite{Rice2010}), Bontempts et al. (\yearcite{Bontemps}), and Vorobyov (\yearcite{Voro2010a}) include the depletion of parent cloud core. 
The effect of depletion is also included in this paper, and its effect on the disk-to-star mass ratio is discussed. 

Time evolution of mass ratio between a central star and a disk with the effect of viscosity is studied 
numerically by several authors. 
\citet{NaNa1994} studied time evolution of young circumstellar disk with viscosity by solving 
the one-dimensional diffusion equation. 
They investigated the disk evolution taking into account of the dynamical accretion onto the disk as well as viscous accretion onto the star. 
They argue that the disk mass can be determined by the competition between mass accretion rate to the disk and to the central star. 
They conclude that the disk mass in the early phase of disk formation tends to be lower than the mass of the star as long as viscous parameter is moderately large ($\alpha~\geq 10^{-2}$). 
However, they considered only the case with the mass accretion rate onto the disk $m_0=0.975$ given by Shu (\yearcite{shu1977}). 
Their conclusion seems to depend on this choice of accretion rate. 
The disk-to-star mass ratio is expected to be different in the case with different accretion rate from the infalling envelope. 

Hueso \& Guillot (\yearcite{HuesoGuillot}) (hereafter HG05) studied the time evolution of the surface density profiles using viscous disk model in order to find compatible set of parameters for the observed T-Tauri stars. 
Visser et al. (\yearcite{Visser}) used one-dimensional viscous evolution model of the disk in order to analyze the chemical processes during the collapse of a molecular cloud core. 
Although time evolution of mass in the star and the disk are also presented in these two studies, 
they assume $m_0=0.975$ as the mass accretion rate onto the disk. 
Thus, the model with the different $m_0$ from Shu's self-similar solution is needed to be studied. 

In this paper, in order to study the disk-to-star mass ratio and the origin of the surface density distribution, 
we investigate the time-dependent evolution of protostellar disks subject to high mass loading from the dynamically collapsing envelope. 
Accretions onto the disk and onto the central star are modeled in \S 2. 
Results of the numerical calculations are shown in \S 3.  
The origin of the surface density profile and disk-to-star mass ratio are discussed in \S 4. 
Summary is given in \S 5. 



\section{Model}
As the simultaneous evolution of the disk and the star, we consider two processes. 
One is infall from the parent cloud core to the circumstellar disk, 
and the other is the accretion from the disk to the central star. 
Mass growth of both the disk and the star is calculated with these two processes. 
Especially, the difference between the mass accretion rate onto the disk and onto the star is taken into account. 
We focus on the accretion phase since transport of angular momentum is inefficient during the runaway collapse phase. 
As initial and outer boundary condition, 
the self-similar solution by SH98 at $t=0$ is adopted. 
The solution by SH98 describes both the runaway collapse and the accretion phase divided by $t=0$. 
In $t>0$, rotation supported disk exists in $r<r_{\rm sh}$ and infalling envelope exists at $r>r_{\rm sh}$ where $r_{\rm sh}$ is a radius of accretion shock. 
In $r<r_{\rm sh}$, the gravitational force and the centrifugal force balance each other with minor contribution from the pressure gradient force. 
Since angular momentum transport is expected to be important in $t>0$ and $r<r_{\rm sh}$, 
we calculate the viscous evolution of a disk in this regime assuming that the disk is geometrically thin. 
\subsection{Accretion from the Infalling Envelope to the Disk}
According to SH98 solution, there is a dynamically infalling flow onto the disk from the envelope. 
Distribution of specific angular momentum $j$ is given by 
\begin{equation}
j(=rv_{\phi})=\omega\frac{GM_{\rm r}}{c_{\rm s}}, 
\label{j}
\end{equation} 
where $M_{\rm r}$ is mass inside of radius $r$. 
A coefficient $\omega$ is a parameter for rotation speed, 
and it is typically $0.2\sim0.4$ at the end of runaway collapse (Matsumoto et al. \yearcite{Matsu}). 
The profile of surface density $\Sigma$ in $r<r_{\rm sh}$ is given by 
\begin{equation}
\Sigma=\Sigma_{1}r^{-1}~({\rm i.e.},~M_{\rm r}\propto r).
\label{initialSigma}
\end{equation}
During the accretion phase, an accretion shock front expands as $r_{\rm sh}=Ac_{\rm s}t$, 
where coefficient $A$ depends on $\omega$ as in Table 1 of SH98. 
The mass accretion rate onto the disk at $r_{\rm sh}$ is written as 
$\dot{M}_{\rm SH}=m_0 c_{\rm s}^3/G$ where $m_0=6$--$10$ for $\omega=0.2$--$0.4$ (Table 1 of SH98). 
Namely, the mass of the disk is written as $M_{\rm disk}(t)=\dot{M}_{\rm SH}t$. 
Above profiles are valid only for the inviscid case where the mass accretion onto the star from the disk is neglected. 
In this paper by assuming that viscous evolution is slower than that of shock propagation, 
as an initial state of numerical calculations, profile given by equation (\ref{initialSigma}) is used for all the radius and 
we define the disk as the region $r<r_{\rm sh}$ as. 
Although we assume that the mass accretion rate onto the shock front is given by $\dot{M}_{\rm SH}$ as in SH98, 
on longer timescales, the mass accretion rate $\dot{M}_{\rm disk}$ onto the disk is expected to decline with time owing to the depletion of the parent cloud core, different from the self-similar solution. 
This will be implemented in \S 2.3. 

\subsection{Accretion from the Disk onto the Star}
With viscosity, the disk loses its mass via viscous accretion onto the central star and becomes different from the inviscid solution (\ref{initialSigma}). 
In this study, we focus on this accretion process 
by assuming that the circumstellar disk is axisymmetric and geometrically thin. 
All the equations are integrated in the vertical direction in cylindrical coordinates ($r,\phi,z$). 
As a result, the system is described by one-dimensional time dependent equation with $t$ and $r$. 
We assume that the pressure gradient force is negligible 
and the accretion speed within the disk is sufficiently slow as 
\begin{equation}
\frac{c_{\rm s}^2}{\Sigma}\frac{\partial \Sigma}{\partial r}\sim 0~{\rm and}~|v_{\rm r}|<<v_{\phi},
\label{rotation}
\end{equation}
where $\Sigma$ is the surface density, $v_{\rm \phi}$ is azimuthal velocity, 
and $v_{\rm r}$ is the radial velocity. 
Then the orbital motions of the disk are due entirely to equating the centrifugal force with gravitational force. 
As a result, viscous evolution of the surface density $\Sigma$ is given by 
\begin{equation}
\frac{\partial \Sigma}{\partial t}=-\frac{1}{r}\frac{\partial}{\partial r}\left[\frac{1}{\frac{\partial(rv_{\phi})}{\partial r}}\frac{\partial}{\partial r}(\nu \Sigma r^3\frac{\partial \Omega}{\partial r})\right],
\label{diffusiveeq}
\end{equation}
where $\Omega=v_{\phi}/r$ is the angular velocity and $\nu$ is viscous coefficient (e.g., \S 7.2 in Hartmann (2008)). 
The original mass conservation equation is given by 
\begin{equation}
\frac{\partial \Sigma}{\partial t}-\frac{1}{r}\frac{\partial}{\partial r}\left(\frac{\dot{M}(r)}{2\pi}\right)=0,
\label{continum}
\end{equation}
where $\dot{M}(r)\equiv 2\pi r \Sigma (-v_{\rm r})$ is mass accretion rate. 
Using equations (\ref{diffusiveeq}) and (\ref{continum}), another form of mass accretion rate is given as 
$\dot{M}(r)=-\frac{2\pi}{\frac{\partial(rv_{\phi})}{\partial r}}\frac{\partial}{\partial r}(\nu \Sigma r^3\frac{\partial \Omega}{\partial r})$, 
and radial velocity $v_{\rm r}$ is given by 
\begin{equation}
v_{\rm r}(r)=-\frac{\dot{M}(r)}{2\pi r \Sigma}=\frac{1}{r\Sigma \frac{\partial(rv_{\phi})}{\partial r}}\frac{\partial}{\partial r}(\nu \Sigma r^3\frac{\partial \Omega}{\partial r}).
\label{vr}
\end{equation}

Hydrostatic equilibrium in $z$-direction is assumed. 
Then volume density is written as 
$\rho(z,r,t)=\rho_{\rm c}(r) {\rm exp} ({-z^2/2{h(r)}^2})$ 
where $h(r)$ is the disk scale height and $\rho_{{\rm c}}(r)$ is the density at $z=0$. 
The relation between $\rho$, $h=c_{\rm s}/\Omega$, and $\Sigma$ are given as $\Sigma=2\rho_{\rm c} h$. 

As angular momentum transport, the standard $\alpha$ prescription for viscous disk (\cite{Pringle}, \cite{Shakura}) is adopted. 
By this prescription, viscous coefficient $\nu$ in equation (\ref{diffusiveeq}) is given by $\nu =\alpha c_{\rm s}h$. 
As for $\alpha$ parameter, we assume
\begin{equation}
\alpha={\rm max}[\alpha _0,\alpha _{\rm G}], 
\label{alpha}
\end{equation}
where $\alpha _0$ is a constant parameter. 
The range of $\alpha _0$ between $10^{-2}$ to $1$ is considered in this paper. 
Angular momentum transport induced by gravitational instabilities is treated in a viscous framework using $\alpha _{\rm G}$ in our model as 
\begin{equation}
\alpha _{\rm G} =\eta(\frac{Q_{\rm c}^2}{Q^2}-1), 
\end{equation}
where $Q_{\rm c}$ is the critical value of Toomre's Q-value (\cite{ToomreQ}) and 
$\eta~(<1)$ is the dimensionless number which represents the efficiency of angular momentum transport (Lin \& Pringle \yearcite{LP90}). 
According to the three-dimensional hydrodynamical calculation in Kratter et al. (\yearcite{Kratter}), 
$\alpha _{\rm G} \sim 1$ is indicated when the disk is gravitationally unstable. 
Boley et al. (\yearcite{Boley}) indicate that angular momentum transport by gravitational torque is important when $Q\leq 1.4$. 
In this paper, $Q_{\rm c}=1.4$ and $\eta=0.2$ are adopted. 

Since the mass of the disk is not negligible to the mass of the central star at $t\sim0$, gravitational force from the disk itself is considered. 
As the simplest approximation, 
gravitational force per unit mass in the radial direction $F_{\rm r}(r)$ is approximated as
\begin{equation}
F_{\rm r}(r)=\frac{GM(r)}{r^2}~{\rm with}~M(r)=M_*\theta(r) +M_{{\rm disk}}(r)~{\rm and}~M_{\rm disk}(r)=\displaystyle \lim_{\epsilon \to 0}\int^r_\epsilon 2\pi r \Sigma(r) dr,
\label{gravi}
\end{equation}
where $M_*$ is mass of the central star, and $M_{\rm disk}(r)$ is the disk mass within $r$. 

Gas in the disk is assumed to be sufficiently cold and to rotate in Keplerian velocity with 
\begin{equation}
v_{\phi}=\sqrt{\frac{GM(r)}{r}}.
\label{kepler}
\end{equation}
Mass of the central star $M_*$ is set to be zero at $t=0$. 
From equations (\ref{initialSigma}) and (\ref{gravi}), we see $M(r)\propto r$ at $t=0$. 
Then equation (\ref{j}) indicates $j\propto r$. 
To treat non-isothermality in high density region, in this paper, 
several cases of temperature distribution are assumed and results are compared. 
As the first case, simple barotropic equation of state is assumed as 
\begin{equation}
c_{\rm s}=
\left\{
\begin{array}{c}
0.2 \quad ({\rm km}/{\rm s}) \quad (\rho <\rho_{\rm crit})\\
0.2 \quad (\frac{\rho}{\rho_{\rm crit}})^{\frac{\gamma-1}{2}} \quad ({\rm km}/{\rm s}) \quad (\rho >\rho_{\rm crit}), 
\label{soundspeedbaro}
\end{array}
\right.
\end{equation}
where $\rho_{\rm crit}=5.0\times10^{-14}~{\rm g}/{\rm cm}^2$ is the critical density above which the equation of state becomes non-isothermal 
with specific heat ratio $\gamma=1.4$. 
This profile is inspired by the result of spherically symmetric radiation hydrodynamical calculation (\cite{MasunagaInutsuka2000}). 

As the second case, the radial distribution of the sound speed in a disk is given by 
\begin{equation}
c_{\rm s}=
\left\{
\begin{array}{c}
0.20~(\rm km/\rm s) \quad (r>100~\rm AU)\\
c_{{\rm s},0}(\frac{T}{T_0})^{-\frac{1}{2}}\propto
\left\{
\begin{array}{c}
0.20~(\frac{r}{100\rm AU})^{-\frac{1}{4}}~(\rm km/\rm s) \quad (r<100~{\rm AU}) \quad (\rm flared\;disks )\\
0.20~(\frac{r}{100\rm AU})^{-\frac{3}{8}}~(\rm km/\rm s) \quad (r<100~{\rm AU}) \quad (\rm flat\;disks).\\
\end{array}
\right.
\label{soundspeedirrad}
\end{array}
\right.
\end{equation}
This case corresponds to a disk heated by the radiation from the central star. 
In this model, only the inner region ($r \leq 100~{\rm AU}$) is assumed to be heated by stellar irradiation.  
As the temperature distributions, we consider the flared disk and the flat disk. 
Sound speed $c_{{\rm s},0}$ and temperature $T_0$ are assumed to satisfy the condition 
$c_{\rm s}=0.20~{\rm km}/{\rm s}$ at $r=100~{\rm AU}$ in each case. 

\subsection{Initial and Boundary Conditions}
We solve equation (\ref{diffusiveeq}) numerically to calculate time evolution of $\Sigma(r)$ in $5<r<1790~\rm AU$. 
Then $\dot{M}(r)$, $M_*$, $M_{\rm disk}(r)$, and $v_{\phi}(r)$ are calculated. 
Radial velocity $v_{r}(r)$ is calculated using equation (\ref{vr}). 
Hereafter we assume $\omega=0.3$. 
In this case, mass accretion rate onto the disk is $1.6\times 10^{-5}~{\rm M}_{\odot}/{\rm yr}$. 
With this mass accretion rate, cloud core initially with $1 {\rm M}_{\odot}$ depletes in $\tau=6.3 \times 10^{4}~{\rm yr}$. 
By this time, accretion shock front $r_{\rm sh}=0.676 c_{\rm s}t$ reaches $1790~{\rm AU}$. 

For simplicity, the propagation of the shock front is not solved in this paper. 
Instead, a disk in $5<r<1790~{\rm AU}$ without viscosity is assumed at the initial state. 
This disk, which is SH98 solution at $t\sim 6.3\times 10^{4}~{\rm yr}$ 
approximates the initial condition for the disk growth with viscosity 
in the case with small $\alpha << 1$.

As initial condition of numerical calculation, surface density $\Sigma=\Sigma_1 r^{-1}$ (equation (\ref{initialSigma})) is used in $5<r<1790~\rm AU$. 
The coefficient $\Sigma_1$ in our model is given in Appendix A1, and it is at most $20\%$ larger than the solution in the accretion phase by SH98. 
Difference originates from the pressure gradient force which is ignored in our calculation. 
In Appendix A2, we show that the pressure gradient force is smaller than the gravitational force in $r>1~{\rm AU}$ for the case with $\omega=0.3$. 

As the outer boundary condition, the mass accretion rate $\dot{M}_{\rm disk}(= \dot{M}_{\rm SH})$ from envelope to the disk at $r_{\rm sh}$ is assumed to be temporary constant before the characteristic time $\tau$. 
We regard the radius of the shock front $r_{\rm sh}= Ac_{\rm s}t$ in SH98 
as the outer edge of disk in our model. 
We define the disk mass as the disk mass within $r\leq r_{\rm sh}$. 
The disk mass is written as $M_{\rm disk}=M_{\rm disk}(r=r_{\rm sh})=\int^{r_{\rm sh}}_{0}2\pi r \Sigma dr$. 
Two types of $\dot{M}_{\rm disk}$ are assumed as the boundary condition after the depletion time $\tau$. 
One is the case where a constant dynamical flow onto the disk is assumed even after $\tau$. 
The other is the case where the mass accretion rate $\dot{M}_{\rm disk}$ decreases to zero as
\begin{equation}
\dot{M}_{\rm disk}=
\left\{
\begin{array}{c}
\dot{M}_{\rm SH} \quad ({\rm for}~t<\tau)\\
0 \quad({\rm for}~t>\tau),
\end{array}
\right.
\label{depletion}
\end{equation}
which represents the depletion of the parent cloud core. 

The inner boundary radius is set to be $r_{\rm ib}=5~\rm AU$. 
Mass entered into $r_{\rm ib}$ is added to the system of the central star and the disk within $r_{\rm ib}$. 
Surface density of the disk in $r<r_{{\rm ib}}$ is extrapolated assuming that the radial profile has the same power index as the region $r\gtrsim r_{\rm ib}$. 
Then amount of mass accreted by the central star is calculated by 
\begin{equation}
M_*(t+dt)=M_*(t) \quad + \quad \dot{M}(r=r_{\rm ib})dt \quad -\quad [M_{\rm disk}(r_{\rm ib},t+dt)~-~M_{\rm disk}(r_{\rm ib},t)].
\label{centralstarmass}
\end{equation}

\subsection{Unsteady Evolution of the Isothermal Disk with Constant Mass Flux onto the Disk}
As the benchmark problem, isothermal viscous disk is used to test our numerical calculations. 
Numerical results are compared to self-similar solutions by T99. 
As the time evolution of isothermal disk, 
absolute value of radial velocity $|v_{\rm r}|$, 
azimuthal velocity $v_{\phi}$, 
mass of star $M_*$, and disk $M_{\rm disk}(r)$,
surface density $\Sigma$, and mass accretion rate $\dot{M}$ for the case with $\alpha=\alpha_0=0.1$ and $\omega=0.3$ are calculated. 
In this subsection, effective viscosity by GI $\alpha_{G}$ is set to be zero,  
and the sound speed is fixed to $c_{\rm s}=0.20~{\rm km}/{\rm s}$ in all radius.

\begin{figure}
\begin{center}
\FigureFile(80mm,80mm){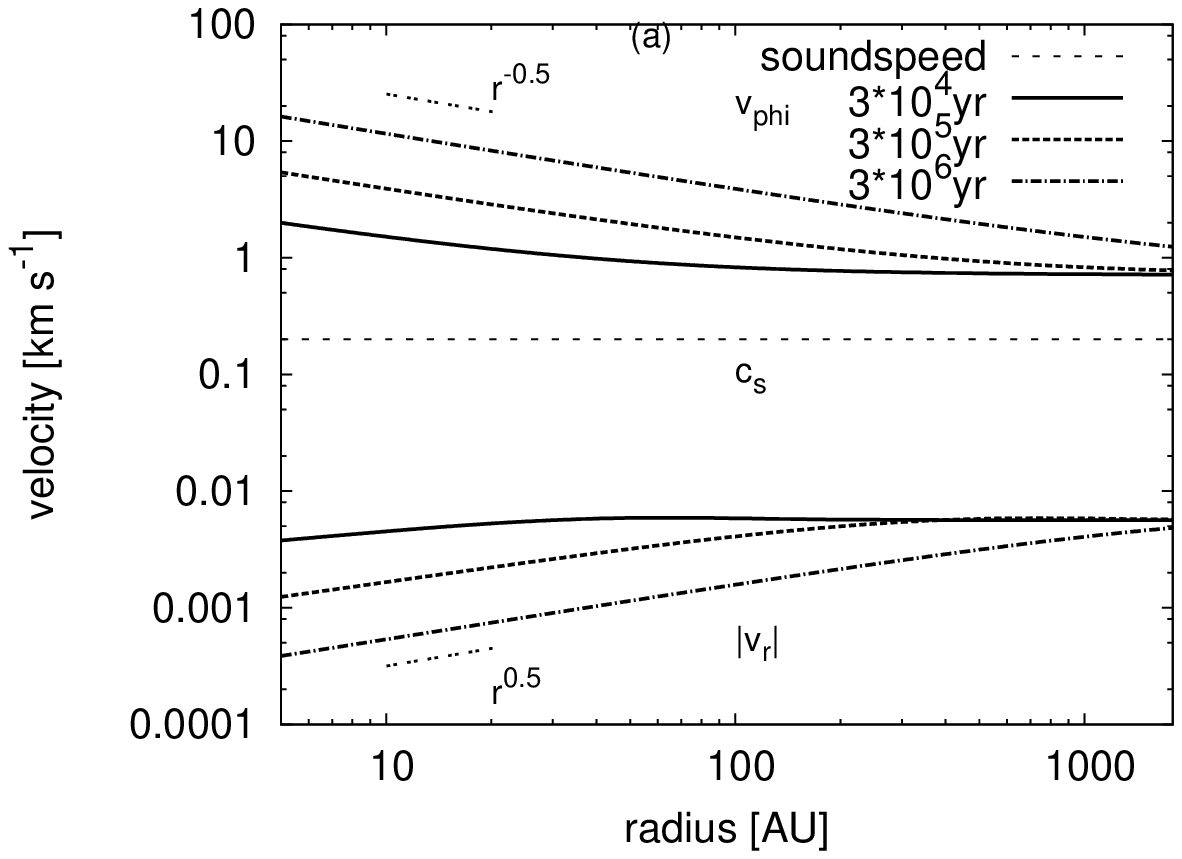}
\FigureFile(80mm,80mm){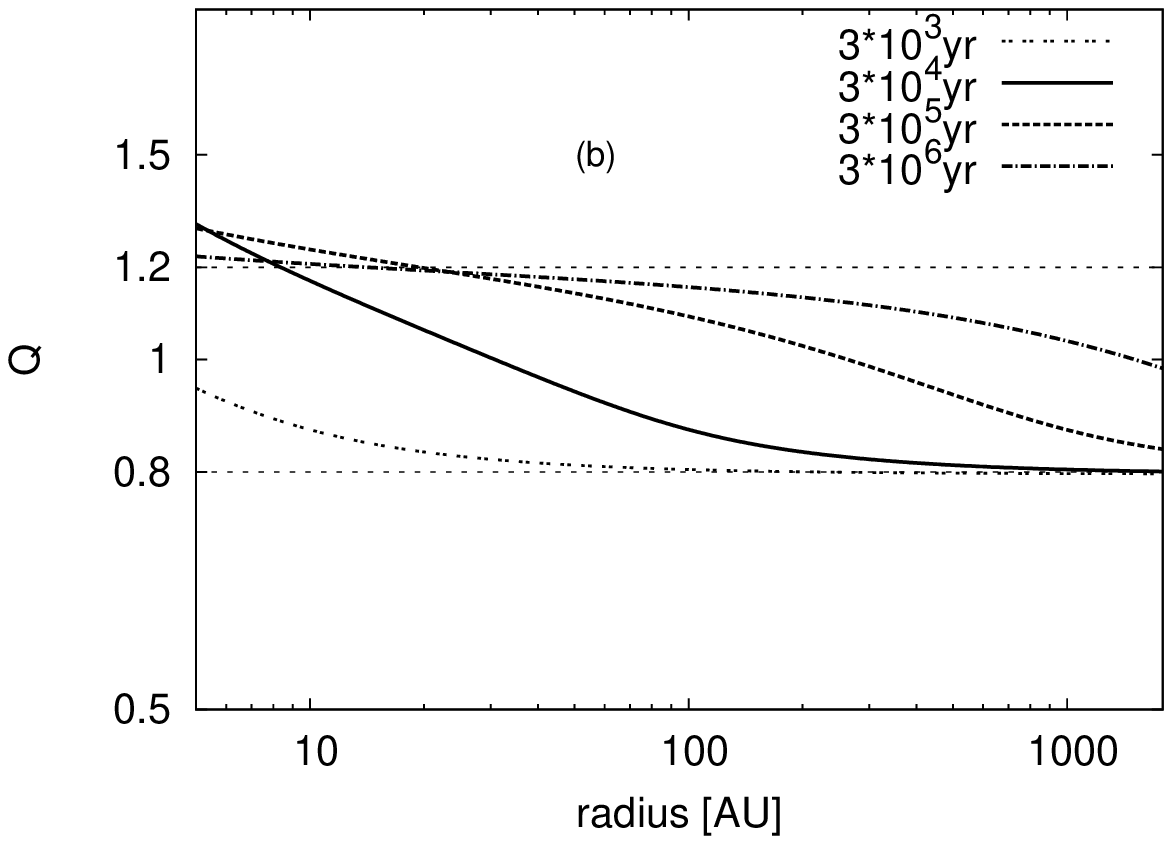}
\FigureFile(80mm,80mm){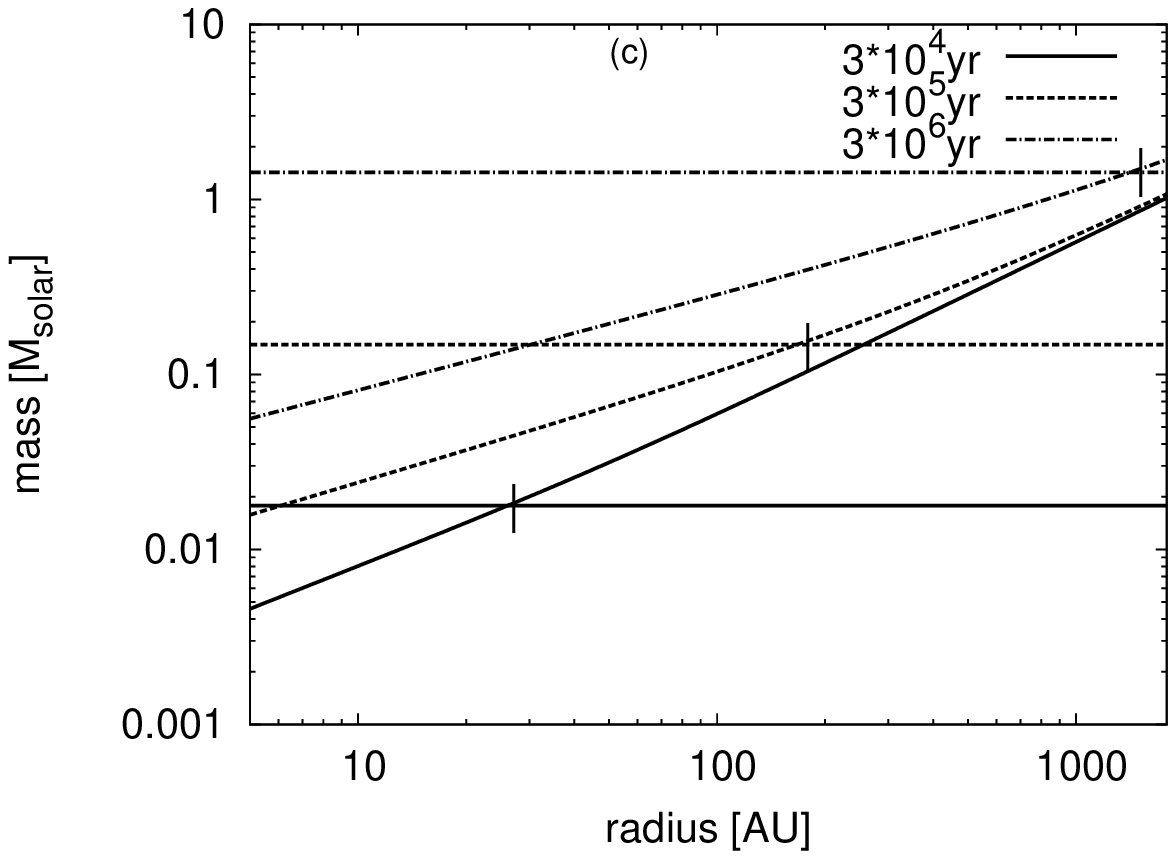}
\FigureFile(80mm,80mm){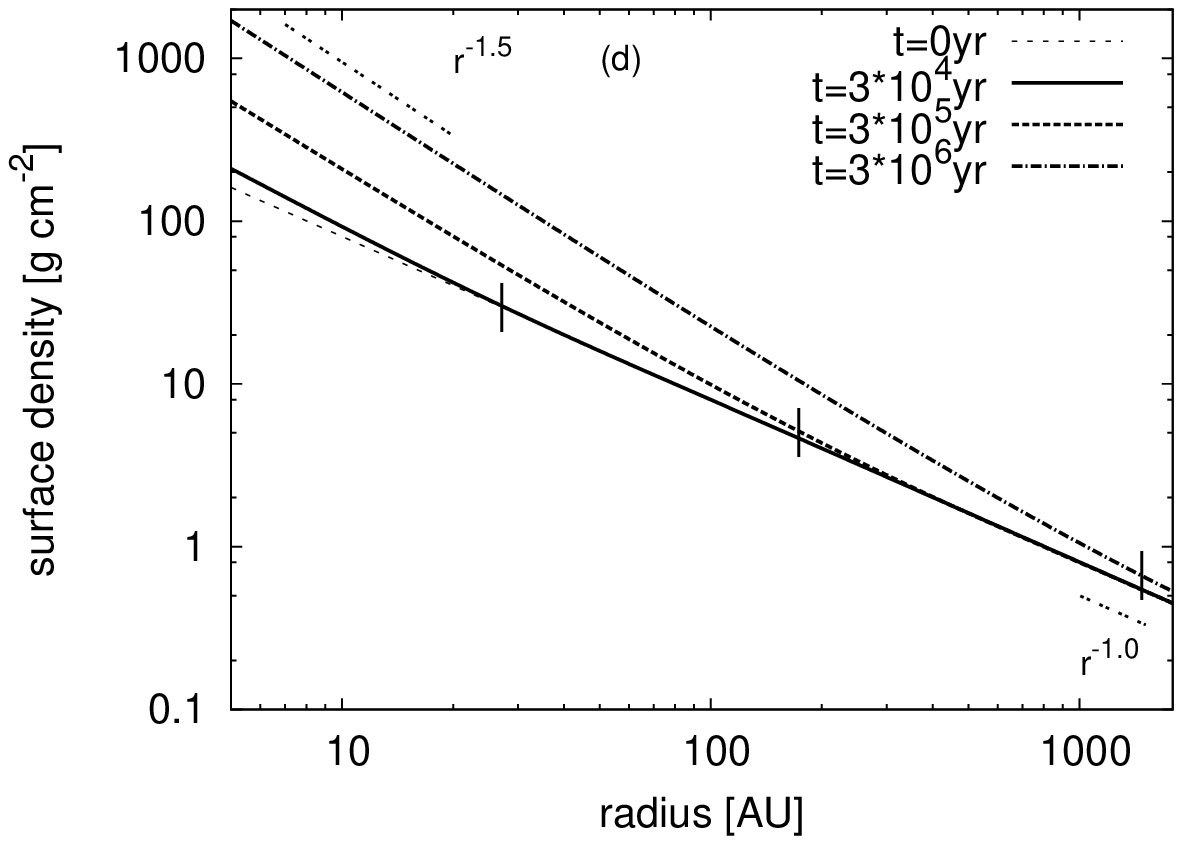}
\FigureFile(80mm,80mm){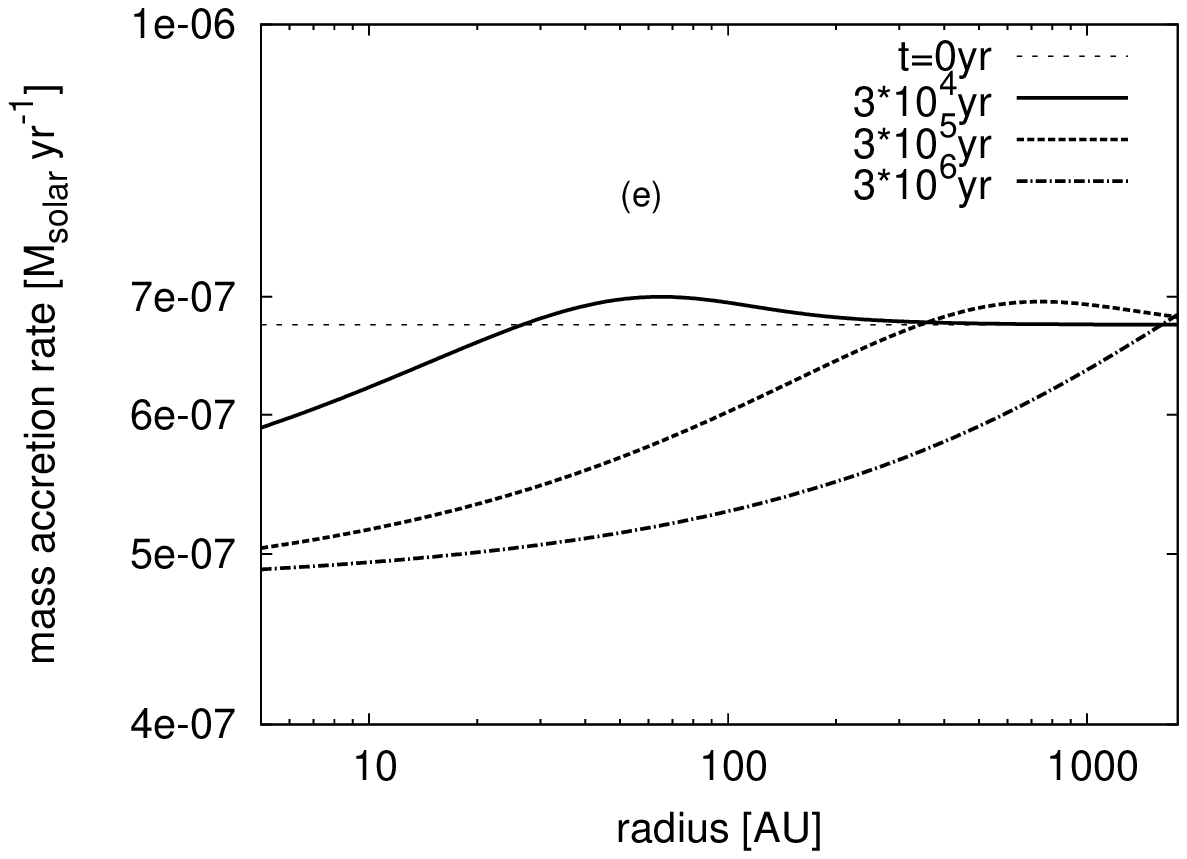}
\end{center}
\caption{Radial profiles of the absolute value of radial velocity $|v_{\rm r}|$, 
azimuthal velocity $v_{\phi}$ (a), Toomre's Q-value (b),
star and disk mass $M_*$, $M_{\rm disk}(r)$ (c),
surface density $\Sigma$ (d), and mass accretion rate $\dot{M}$ (e) for the case with $\alpha=0.1$ and $\omega=0.3$. Isothermal equation of state is assumed.}
\end{figure} 
Velocities $|v_{\rm r}|$ and $v_{\phi}$ are shown in Figure 1 (a). 
It is seen that the assumption $|v_{\rm r}|<<v_{\phi}$ is satisfied in all the time and radius. 
Area with $|v_{\rm r}|\propto r^{0.5}$ and $v_{\phi}\propto r^{-0.5}$ appears from the inner radius and it spreads with time according to the viscous accretion. 
The azimuthal velocity $v_{\phi}\propto r^{-0.5}$ in the inner region represents usual Keplerian rotation velocity around a point mass. 
This indicates the formation  and growth of the central star as a result of viscous effect. 
In the outer region, azimuthal velocity $v_{\phi} \propto r^{0}$ is the same radial dependence as initial state. 
This velocity profile is as a result that the disk mass is larger than the central star, 
where gravitational force has profiles $F_{\rm r}(r) \propto r^{-2}~(r \rightarrow 0)$ and
$F_{\rm r}(r) \propto r^{-1}~(r \rightarrow \infty)$ 
(c.f., Mestel \yearcite{Mestel} and Hayashi et al. \yearcite{Hayashidisk}). 

Furthermore, in the inner region, azimuthal velocity $v_{\phi}$ increases 
and radial velocity $|v_r|$ decreases with time. 
All of these properties of numerical results are confirmed to be consistent with analytic formulae for the time-dependent self-similar solutions for viscous flow given by 
\begin{equation}
v_{\phi}=(\frac{3\alpha c_{\rm s}^3t}{Qr})^{\frac{1}{2}},~v_r=-(\frac{3\alpha Qc_{\rm s}r}{4t})^{\frac{1}{2}}
\label{vinner}
\end{equation} 
in the inner region $r\rightarrow 0$ (c.f., equation (25) in T99) and 
\begin{equation}
v_{\phi}=\frac{2\sqrt{2} c_{\rm s}}{Q},~ v_r=-\frac{\alpha c_{\rm s}Q}{2\sqrt{2}}
\label{vouter}
\end{equation}
in the outer region $r\rightarrow \infty$ (c.f., equations (24) and (34) in T99). 

Toomre's Q-value (\cite{ToomreQ}) $Q$ in equations (\ref{vinner}) and (\ref{vouter}) is given by
\begin{equation}
Q=\frac{c_{\rm s}\kappa}{\pi G \Sigma},~{\rm where}~ \kappa=\Omega\big(4+2\frac{d{\rm log}\Omega}{d{\rm log}r}\big)^{\frac{1}{2}}. 
\end{equation}
Radial profiles of $Q$ in several time epochs are shown in Figure 1 (b). 
It is seen that $Q$ in the inner region increases from the initial value 
$0.8$ to $1.2$ in the long time limit according to the growth of the central star. 

In Figure 1 (c), $M_{\rm disk}(r)$ and $M_*$ for several different time epochs are shown. 
Mass within $r$ is given by $M(r)=M_* \theta(r) + M_{\rm disk}(r)$, 
where the $M_*$ is mass of the central star and 
$M_{\rm disk}(r)$ 
is mass of the disk within $r$. 
Each horizontal line represents the mass of the central star $M_*$. 
Each line increasing with radius represents the disk mass $M_{\rm disk}(r)$ within $r$. 
It is seen that both the disk mass and the central star mass increase with time. 
As the central star mass $M_*$ increases, the radius $r_*(t)$ 
where $M_{\rm disk}(r)$ equals to $M_*$ shifts to larger radius. 
This radius $r_*(t)$ is close to the characteristic radius which divides $v_{\phi}$ into above two branches $v_{\phi}\propto r^{-1/2},~r^{-1}$ (equations \ref{vinner} and \ref{vouter}). 

Figure 1 (d) shows the surface densities $\Sigma(r)$ at several different time epochs. 
Thin dotted line represents the initial state $\Sigma\propto r^{-1}$. 
It is seen that area with $\Sigma \propto r^{-1.5}$ appears from inner radius and it spreads with time. 
In the outer region, the surface density profile keeps its initial distribution $\Sigma\propto r^{-1}$. 
In the inner region, the surface density increases with time. 
The asymptotic formulae of $\Sigma$ are given by
\begin{equation}
\Sigma=\frac{1}{\pi G}(\frac{3\alpha c_{\rm s}^5t}{Q^3r^3})^{\frac{1}{2}}~{\rm for}~r\rightarrow 0 \quad {\rm and}\quad \Sigma=\frac{4c_{\rm s}^2}{\pi GQ^2r}~{\rm for}~r \rightarrow~\infty
\end{equation} 
(c.f., equations (24) and (25) in T99).
The inner branch with $\Sigma \propto r^{-1.5}$ and outer branch with $\Sigma\propto r^{-1}$ connect smoothly near the time dependent radius $r_*(t)$. 

Distributions of mass accretion rate $\dot{M}(r)$ in several time epochs are shown in Figure 1 (e). 
Although $\dot{M}(r)=0$ is set as an initial condition, 
the viscosity induces spatially constant mass accretion rate $\dot{M}(r)= 6.75\times 10^{-7}~{\rm M}_{\odot}/{\rm yr}$ instantaneously at $t\sim 0$. 
As time passes, the amount of $\dot{M}(r)$ decreases into 0.70 times in a long time limit. 
The time-dependent properties of $\dot{M}(r)$ are given by
\begin{equation}
\dot{M}=\frac{3\alpha c_{\rm s}^3}{QG}~{\rm for}~r\rightarrow 0~{\rm and}~\dot{M}=\frac{2\sqrt{2} \alpha c_{\rm s}^3}{QG}~{\rm for}~r \rightarrow \infty
\end{equation} 
(c.f., equations (25) and (34) in T99). 
The different two asymptotic branches of $\dot{M}(r)$ are connected smoothly near the radius $r_*(t)$.  

In summary, from Figure 1 it is shown that our numerical results for the isothermal disk are consistent with the self-similar solution. 
This assures the accuracy of our result of numerical calculation.  
\section{Results}




\subsection{Unsteady Evolution of Non-isothermal Disks}

\subsubsection{A Disk with Barotropic Equation of State}
\begin{figure}
\begin{center}
\FigureFile(80mm,80mm){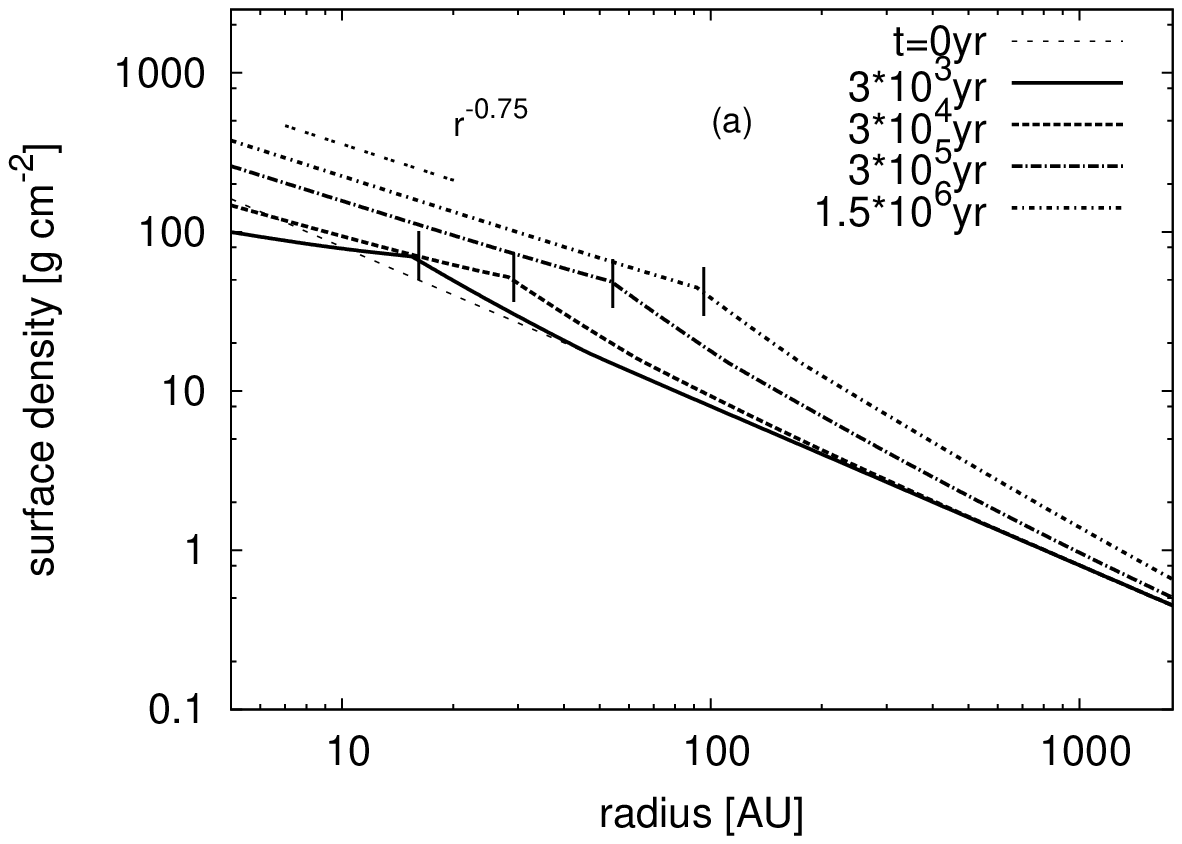}
\FigureFile(80mm,80mm){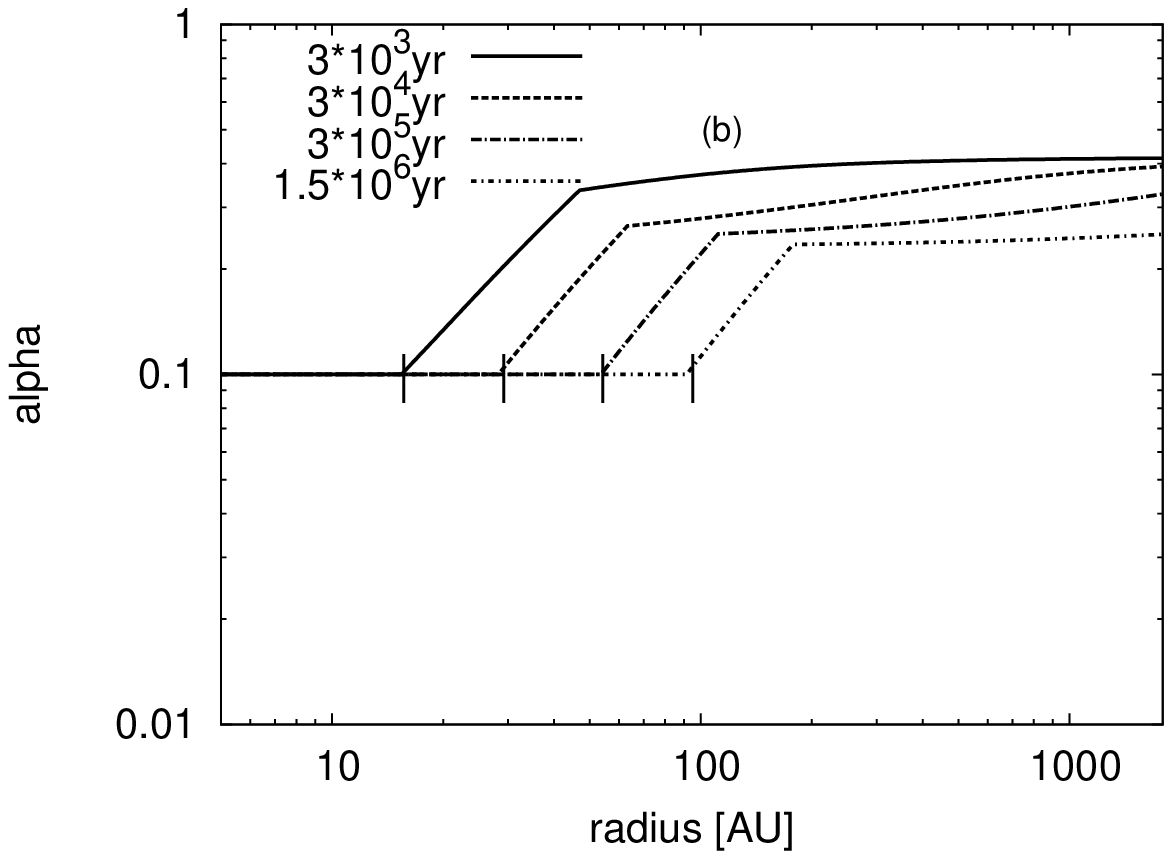}
\FigureFile(80mm,80mm){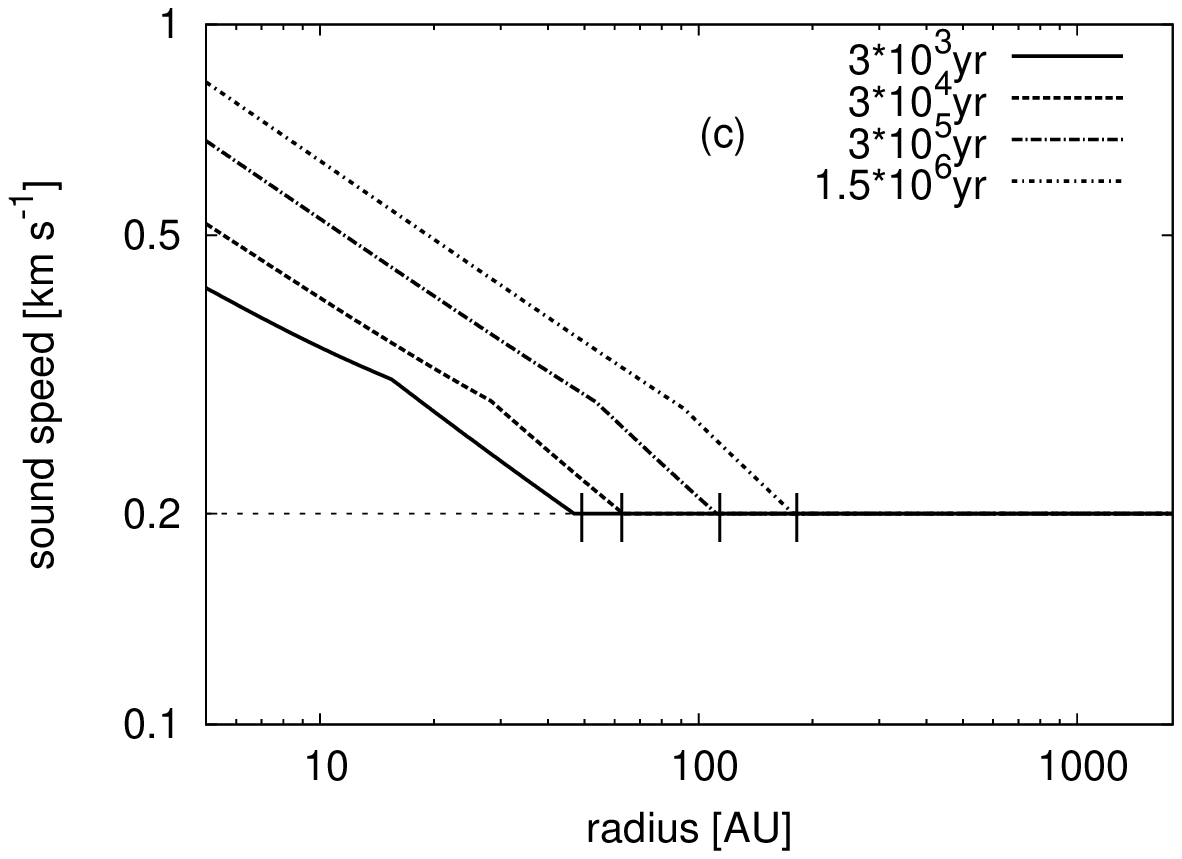}
\FigureFile(80mm,80mm){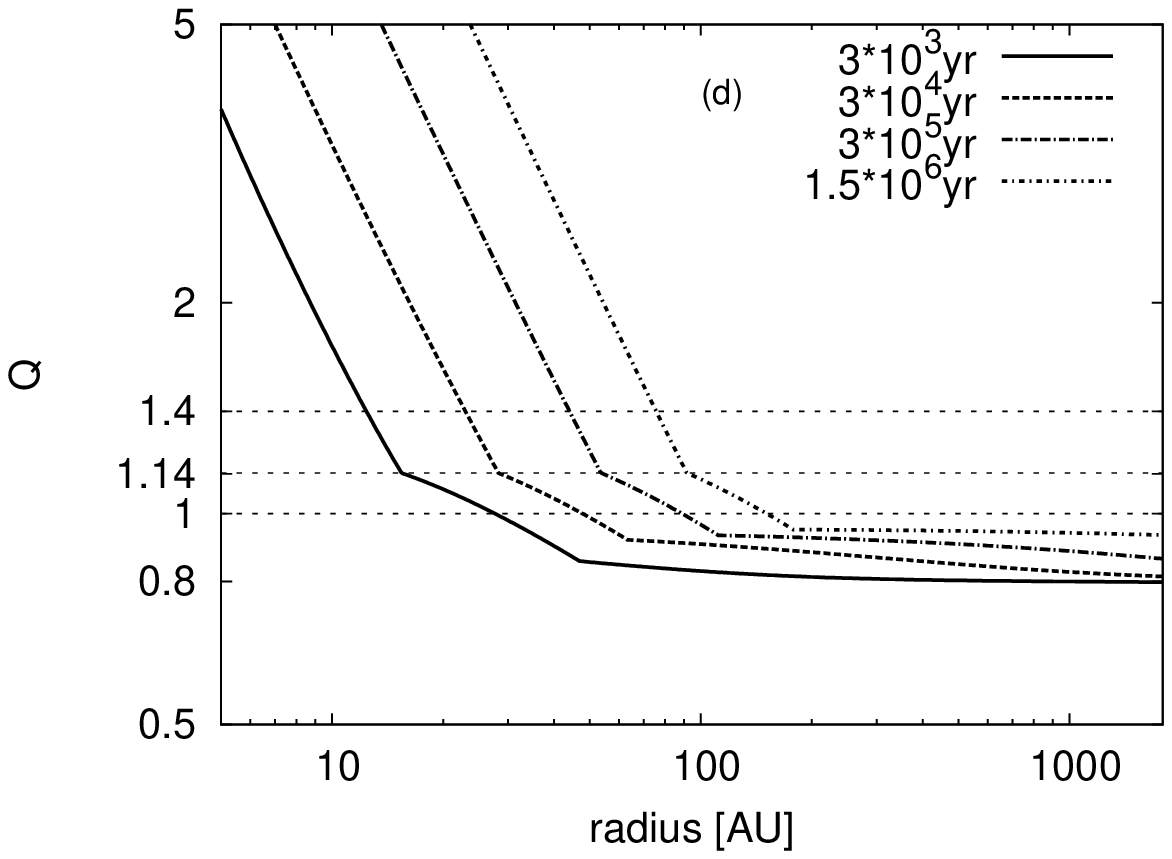}

\caption{Radial profiles of surface density $\Sigma$ (a), effective viscous parameter $\alpha$, (b), sound speed $c_{\rm s}$ (c), and Toomre's Q-value (d) for the case with $\alpha=0.1,~\omega=0.3$, and $\rho_{\rm crit}=5.0\times10^{-14}~{\rm g}/{\rm cm}^2$. 
As the equation of state, equation (\ref{soundspeedbaro}) is assumed.}
\end{center}
\end{figure}
As the first model of non-isothermal disk, radial dependence of sound speed with equation (\ref{soundspeedbaro}) is studied. 
As the time evolution of non-isothermal disk, surface density $\Sigma$, the effective viscous parameter $\alpha$, Toomre's Q-value $Q$, and sound speed $c_{\rm s}$ are shown in Figure 2 for the case with $\rho_{\rm crit}=5.0\times10^{-14}~{\rm g}/{\rm cm}^3$, $\omega=0.3$, and $\alpha_0=0.1$. 
Initial surface density is the same as equation (\ref{initialSigma}). 
At the initial state, there is a non-isothermal region inside of $43~{\rm AU}$. 
In Figure 2 (a) distributions of the surface density $\Sigma$ in several time epochs are shown. 
After the surface density decreases in the inner region during $0<t<3\times 10^3$ yr, 
area with $\Sigma\propto r^{-0.75}$ appears from the inner radius. 
This radial profile is different from $\Sigma\propto r^{-1.5}$ in the isothermal disk in \S 2.4. 
The inner branch with $\Sigma\propto r^{-0.75}$ is connected to other region at the critical radius $r_{\rm crit}$ 
which are $15,~28,~53,~{\rm and}~92 ~{\rm AU}$ at $3\times 10^3,~3\times 10^4,~3\times 10^5,~{\rm and}~1.5\times 10^6~\rm yr$, respectively. 
Radial distributions of parameter $\alpha$ in several time epochs are shown in Figure 2 (b). 
In Figure 2 (b), it is seen that profile of $\alpha$ is spatially constant inside a characteristic radius. 
We call this radius $r_{\rm visc}$ 
which are $15,~28,~54,~{\rm and}~91 ~{\rm AU}$ at $3\times 10^3,~3\times 10^4,~3\times 10^5,~{\rm and}~1.5\times 10^6~\rm yr$, respectively. 
By comparing Figure 2 (a) and 2 (b), it is seen that the critical radius $r_{\rm crit}$ is near the radius $r_{\rm visc}$. 
The radial profiles of sound speed in several time epochs are shown in Figure 2 (c).
Figure 2 (a), 2 (b) and 2 (c) indicate the characteristic radius $r_{\rm temp}$, 
which are $46,~62,~{\rm and}~102 {\rm AU}$ at $3\times 10^3,~3\times 10^4,~{\rm and}~3\times 10^5$ yr, respectively. 
Radial profiles of $Q$ in several time epochs are shown in Figure 2 (d). 
From Figure 2 (b) and 2 (d), it is seen that 
effective viscous parameter $\alpha_{\rm G}$ is larger than $\alpha_0=0.1$ when $Q<1.14$. 
Therefore, the critical radius $r_{\rm visc}$ coincides with the radius where $Q=1.14$. 

In summary, from Figure 2 it is shown that the radial profile of the surface density depends on two characteristic radii. 
One is $r_{\rm visc}(t)$ outside which the viscous parameter $\alpha$ is not constant 
and the other is $r_{\rm temp}$ inside which the disk is non-isothermal. 

\subsubsection{Disks Heated by the Irradiation of the Forming Star}
As another model of non-isothermal disk, 
alternative radial dependence of sound speed with equation (\ref{soundspeedirrad}) which corresponds to a flared disk or a flat disk is also studied. 
\begin{figure}
\begin{center}
\FigureFile(80mm,80mm){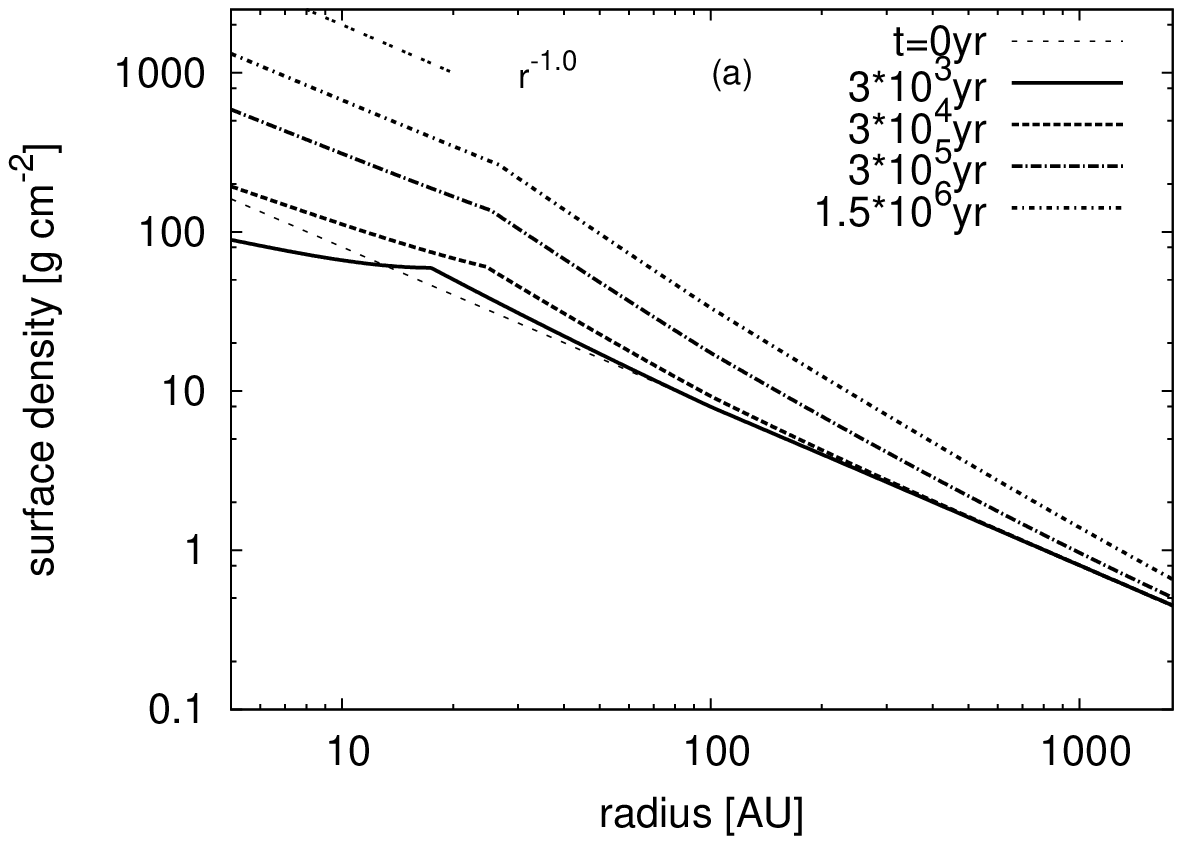}
\FigureFile(80mm,80mm){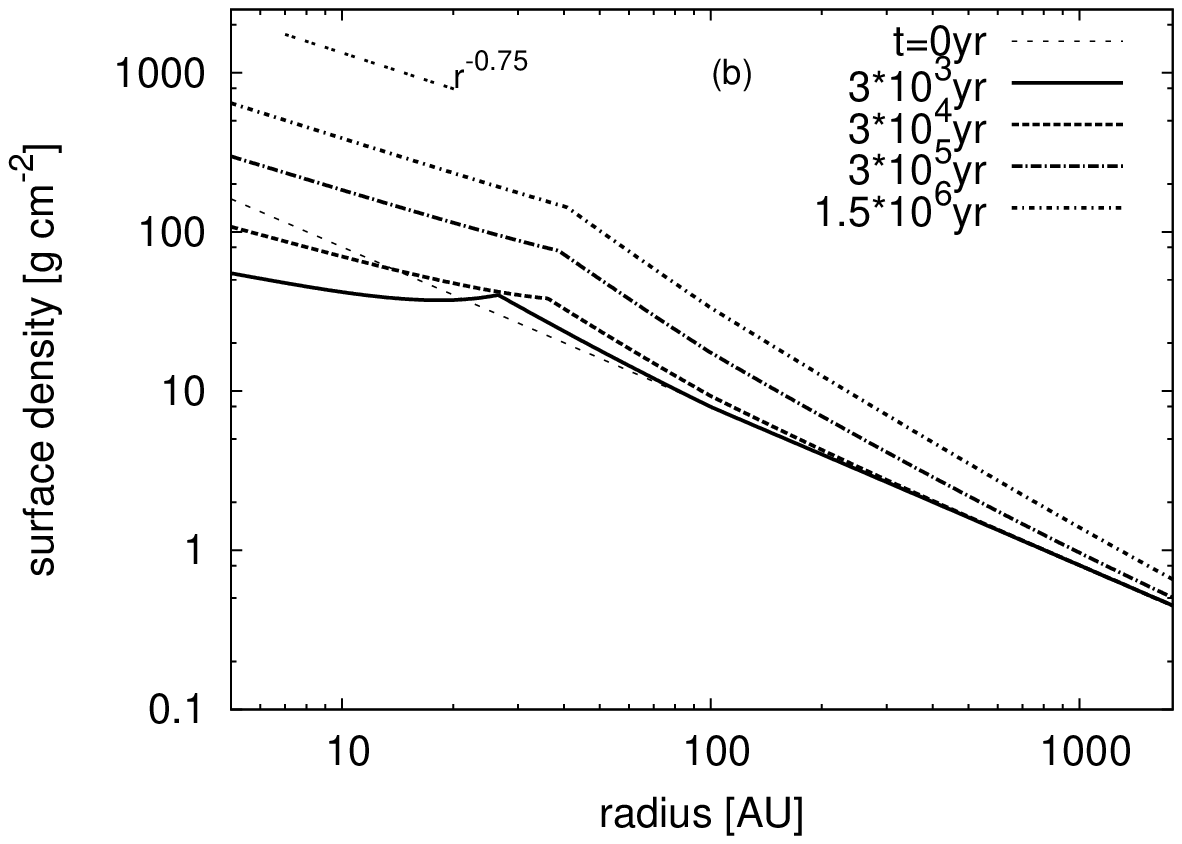}
\FigureFile(80mm,80mm){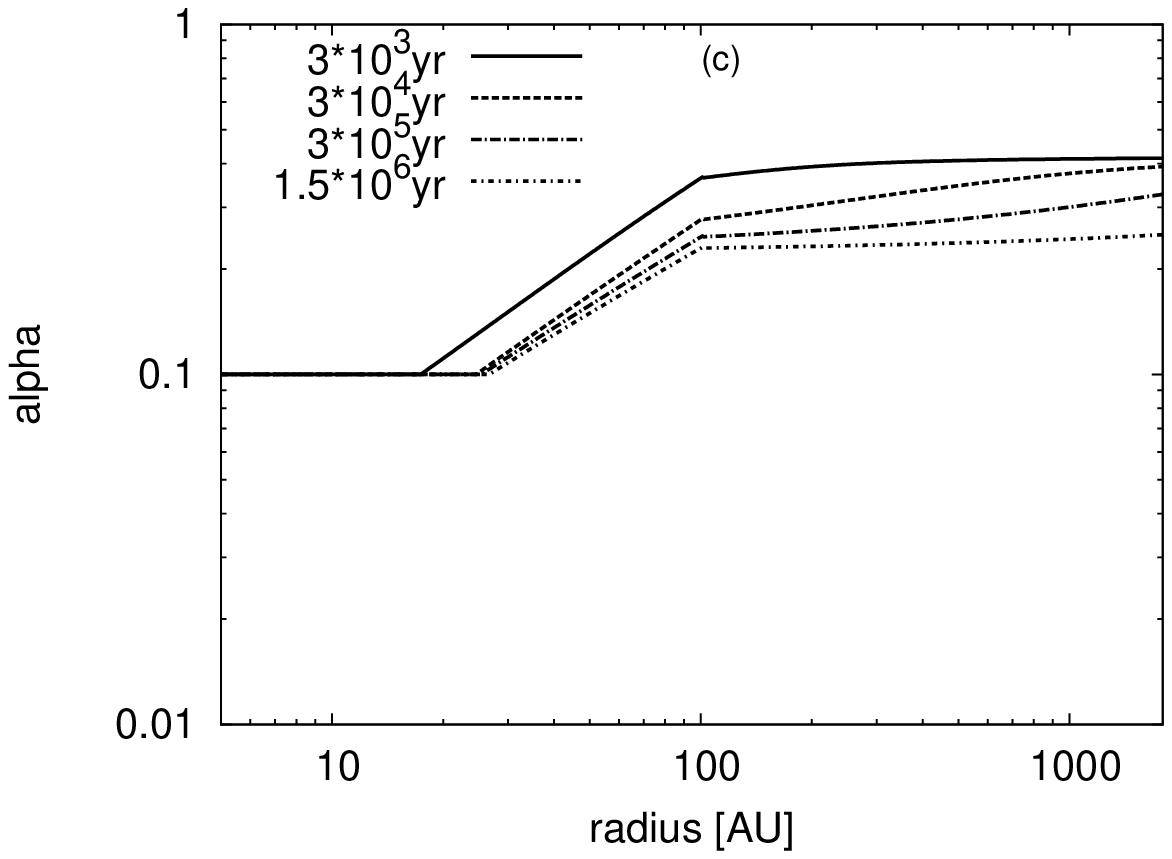}
\FigureFile(80mm,80mm){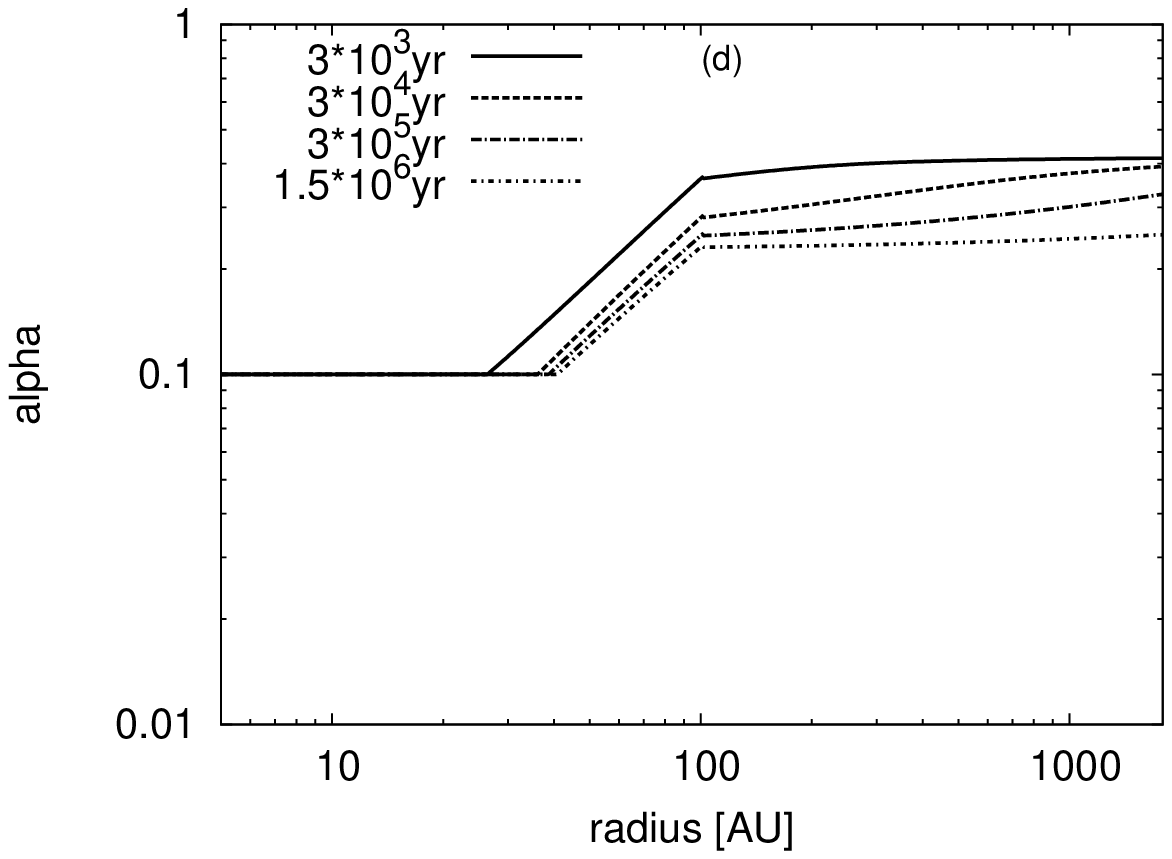}
\caption{Radial profiles of the surface density $\Sigma$ for the flared disk (a), for the flat disk (b), $\alpha$ for the flared disk (c), and for the flat disk (d). Each line represents different time epochs. As the temperature distribution, equation (\ref{soundspeedirrad}) is assumed.}
\end{center}
\end{figure}
The results are shown in Figure 3. 
Distributions of the surface density of the flared disk in several time epochs are shown in Figure 3 (a). 
It is seen that the branch with $\Sigma\propto r^{-1}$ appears from the inner region. 
Distributions of the surface density of the flat disk in several time epochs are shown in Figure 3 (b). 
It is seen that the branch with $\Sigma\propto r^{-0.75}$ appears from the inner region. 
This profile has the same power index as in the case with barotropic equation of state (see \S 4.1 for reason). 

From above results, it is found that the surface density in the inner region depends on the model of temperature profile. 
However, the power index of radial profile of the surface density is within a range between $-1.5$ and $-0.75$ in all the result including previous sections. 
From this fact, we stress that disk with the surface density steeper than $\Sigma \propto r^{-2}$ is not expected to form as long as parameters in this paper are assumed. 
Note that KI02 argued that protoplanetary disk with $\beta > 2$ and $\beta<2 $ in $\Sigma \propto r^{-\beta}$ has qualitatively different property in the view point of planet formation. 

\subsection{Time Evolution of the Disk-to-Star Mass Ratio}
\subsubsection{Case with Constant Mass Flux}
\begin{figure}
\begin{center}
\FigureFile(80mm,80mm){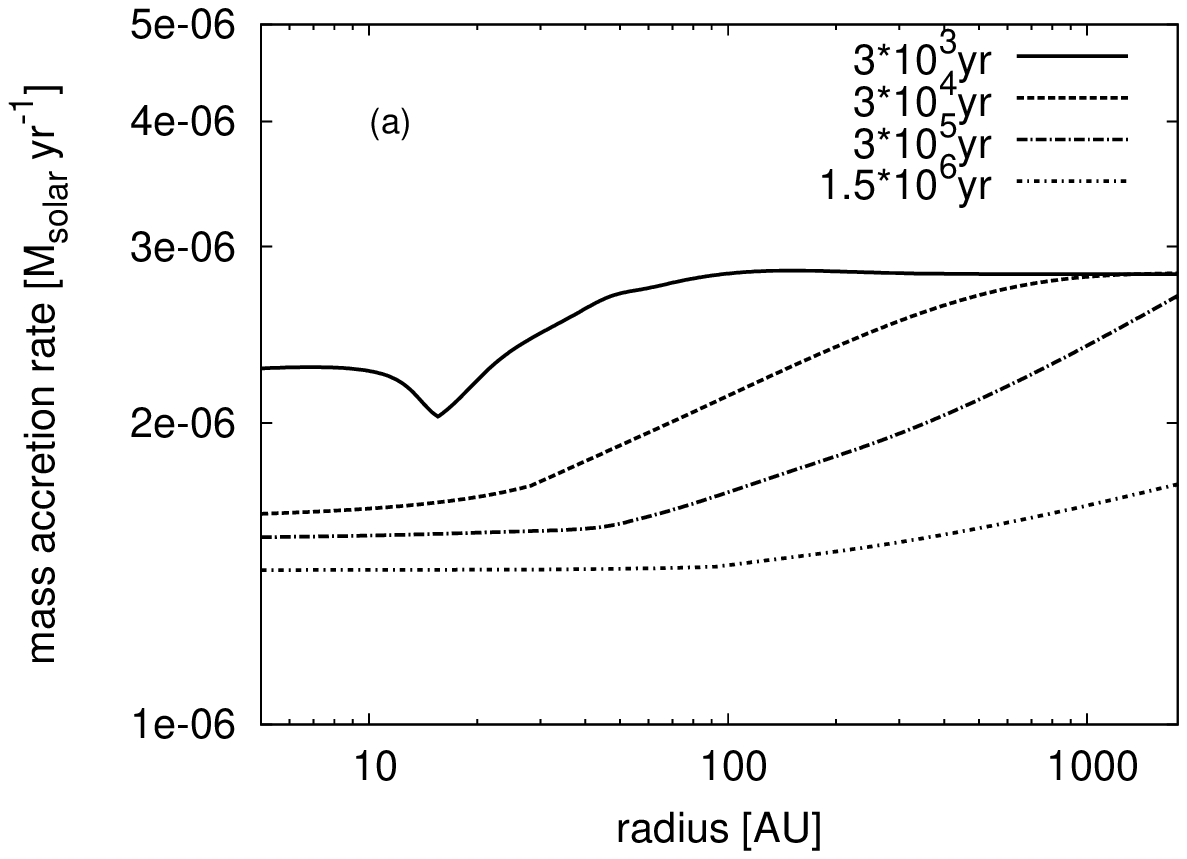}
\FigureFile(80mm,80mm){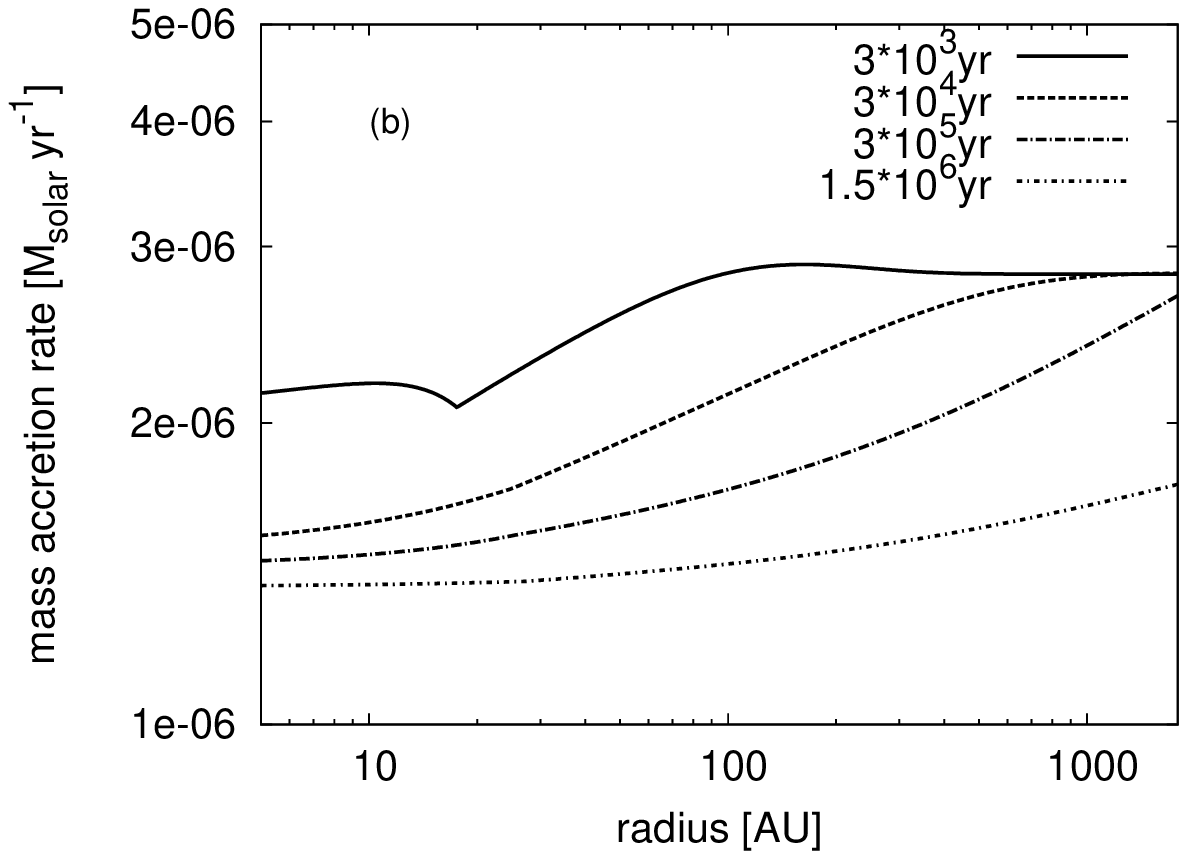}
\caption{Mass accretion rate for the case with barotropic equation of state (a), and for the case with flared disk (b). In each diagram, $\alpha_0=0.1$ is assumed.}
\end{center}
\end{figure}
In this subsection 3.2.1, numerical results are shown for the case where the mass accretion rate onto the disk $\dot{M}_{\rm disk}$ is temporary constant $1.6\times 10^{-5}~{\rm M}_{\odot}/{\rm yr}$ with $\omega =0.3$. 

Radial distributions of mass accretion rate $\dot{M}(r)$ in several time epochs are shown in Figure 4. 
Figure 4 (a) and 4 (b) are for the case with barotropic equation of state and with the flared disk, respectively. 
In both Figure 4 (a) and 4 (b), a region with spatially constant $\dot{M}(r)$ appears in the inner radius and expands with time. 
We regard this region with spatially constant $\dot{M}(r)$ as the quasi-steady state and discuss analytically in \S 4.1. 
\begin{figure}
\begin{center}
\FigureFile(80mm,80mm){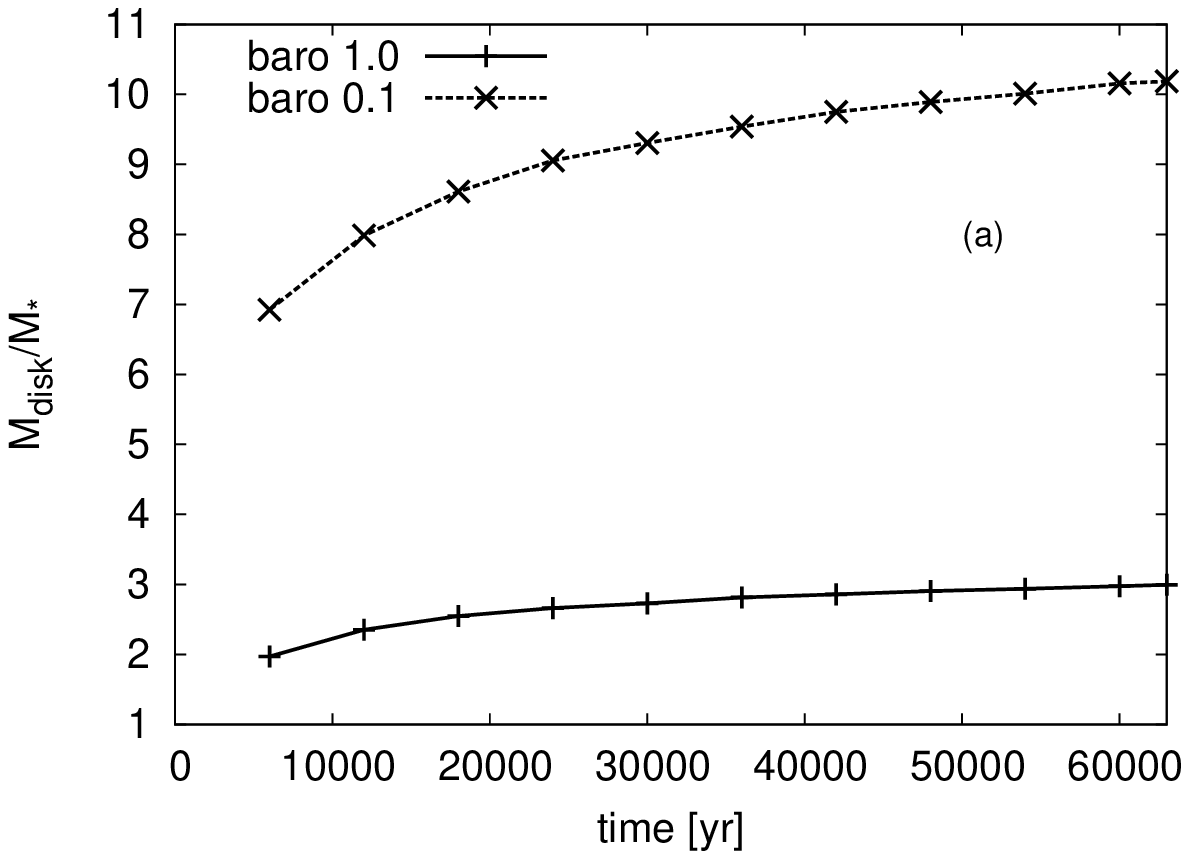}
\FigureFile(80mm,80mm){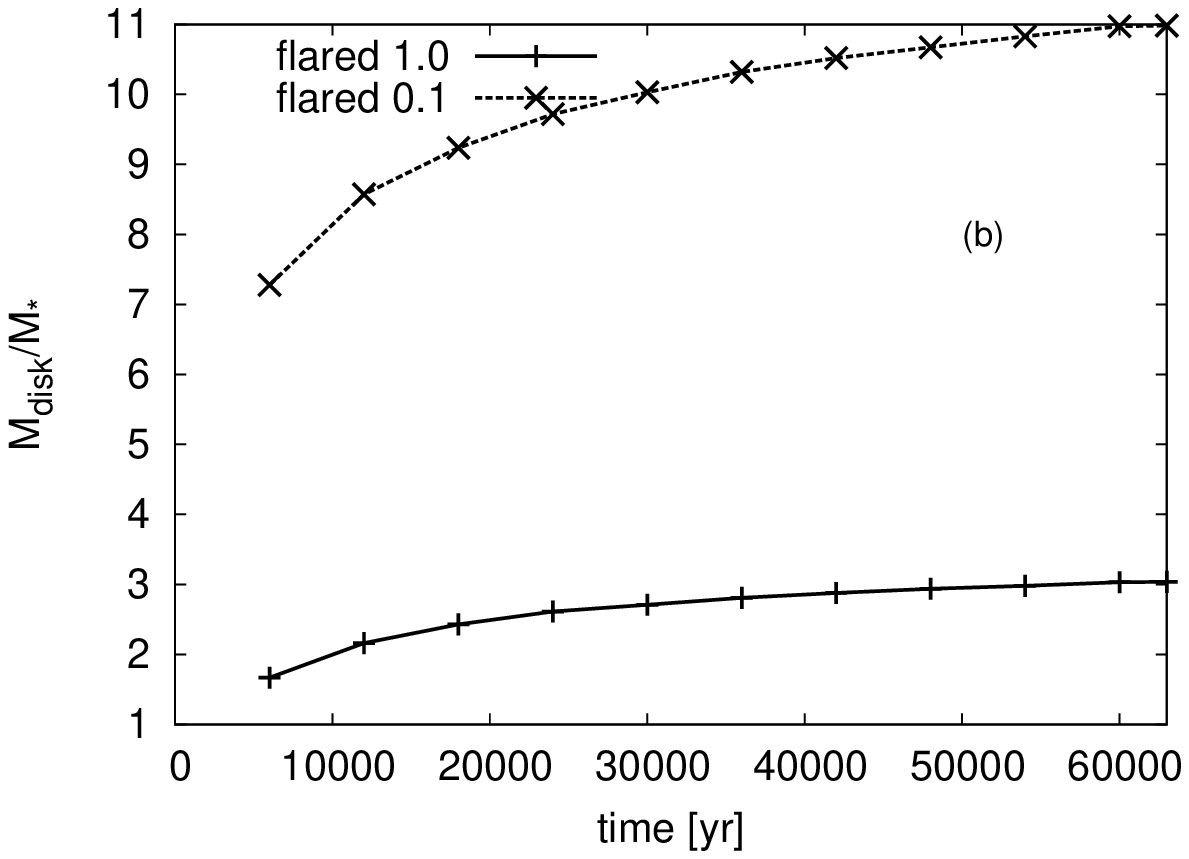}
\caption{Disk-to-star mass ratio for the case with barotropic equation of state (a), and for the case with the flared disk (b). The solid line shows the case with $\alpha_0=1$ and the dotted line shows the case with $\alpha_0=0.1$.}
\end{center}
\end{figure}
In Figure 4 (a), mass accretion rate onto the star $\dot{M}_*$ in the quasi-steady state is approximately $1.5\times 10^{-6}~{\rm M}_{\odot} /{\rm yr}$ for the case with $\alpha_0=0.1$. 
On the other hand, mass accretion rate onto the disk is $1.6\times 10^{-5}~{\rm M}_{\odot} /{\rm yr}$ for the case with $\omega =0.3$ (see \S 2.3). 
Thus, it is clear that mass accretion rate onto the star $\dot{M}_{*}$ is ten times smaller than that onto the disk $\dot{M}_{\rm disk}$. 
This result indicates that the disk-to-star mass ratio is increasing function of time. 
Thus, the disk mass $M_{\rm disk}$ is expected to become larger than mass of the star $M_*$ in the long time limit as long as $\alpha_0<1$ and $\dot{M}_{\rm disk}$ is temporally constant.

In Figure 5, time evolutions of the disk-to-star mass ratio $M_{\rm disk}/M_*$ are shown. 
It is confirmed that the disk-to-star mass ratio is increasing function of time. 
Increasing rate of the disk-to-star mass ratio is larger in the early stage ($t<2\times 10^4~\rm yr$). 
This originates from the fact that mass accretion rate onto the star decreases in the early stage. 
For example, in Figure 4 (a) it is seen that the mass accretion rate inside $r=5~{\rm AU}$ decreases remarkably before $3\times 10^4~\rm yr$. 
This is because initial distribution of the surface density was not for the quasi-steady state. 
After $3\times 10^4~\rm yr$, inner region approaches to the quasi-steady state and mass accretion rate decreases much slowly. 
Furthermore, in Figure 5 it is seen that $M_{\rm disk}/M_*$ is $7$--$10$ for the case with $\alpha_0\sim 0.1$, 
and $2$--$3$ for $\alpha_0 \sim 1$ in $t>6000~\rm yr$. 
This shows that mass of the disk is larger than the central star as long as $\alpha_0<1$ in the long time limit. 
These results are qualitatively same in Figure 5 (a) and 5 (b) with different temperature distribution 
(barotropic equation of state and the flared disk). 

In summary, 
it is found that the disk-to-star mass ratio $M_{\rm disk}/M_*$ tend to be larger than unity as long as $\alpha_0<1$ 
provided that the temporary constant mass flux onto the disk is assumed as SH98. 

\subsubsection{Effect of Decline of the Mass Infall Rate onto the Disk}
\begin{figure}
\begin{center}
\FigureFile(80mm,80mm){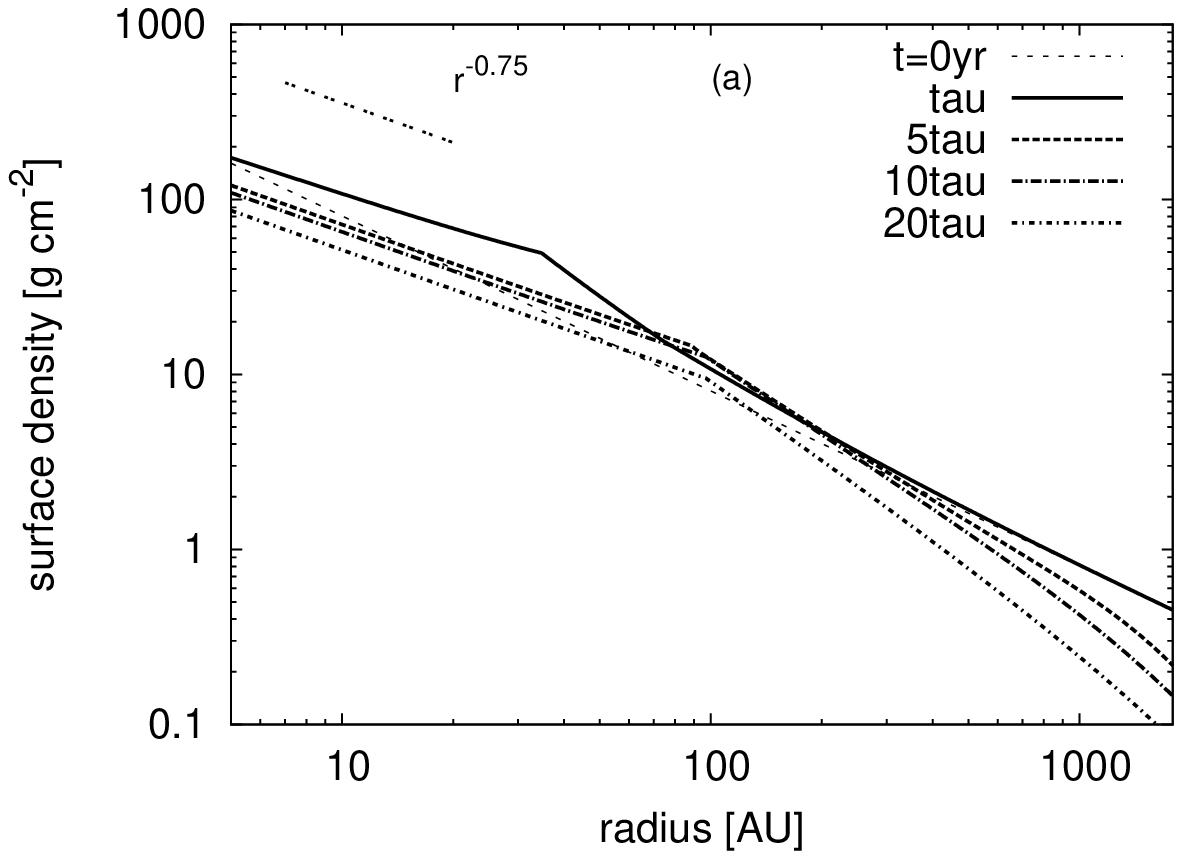}
\FigureFile(80mm,80mm){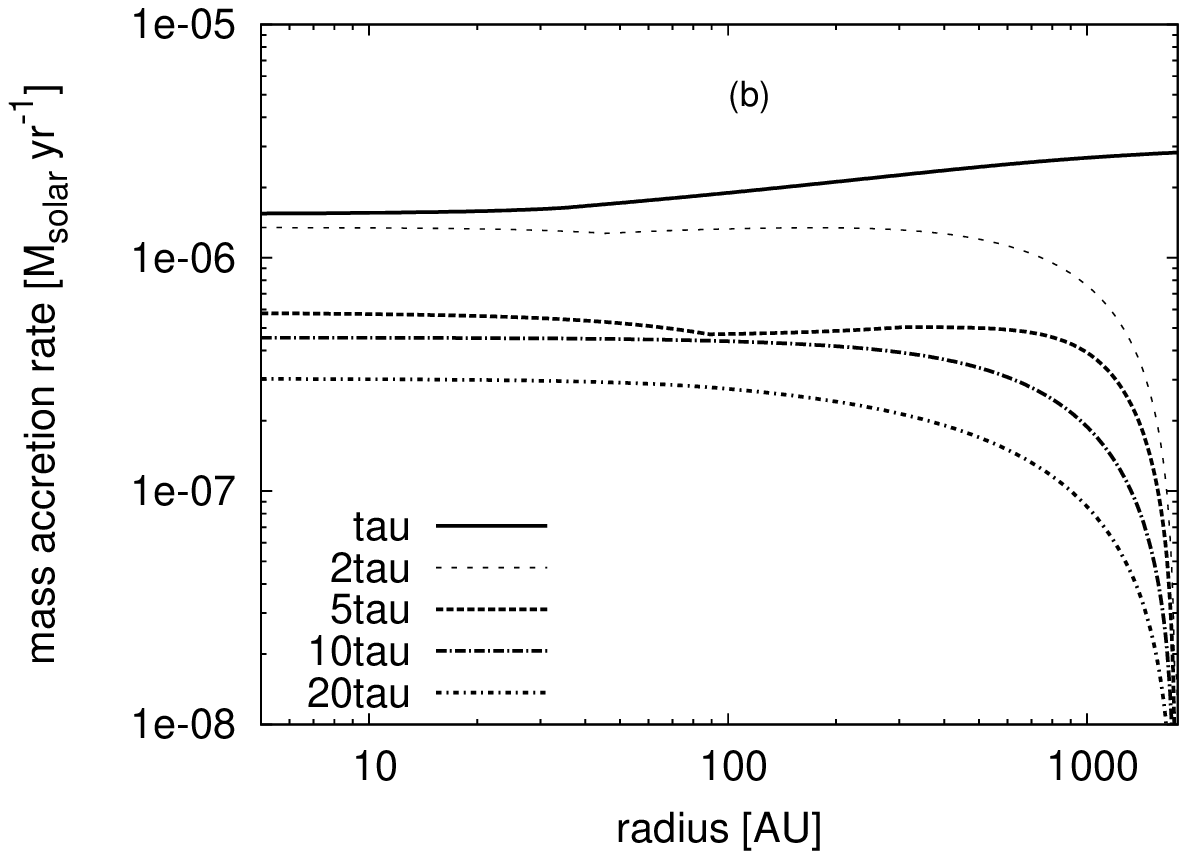}
\FigureFile(80mm,80mm){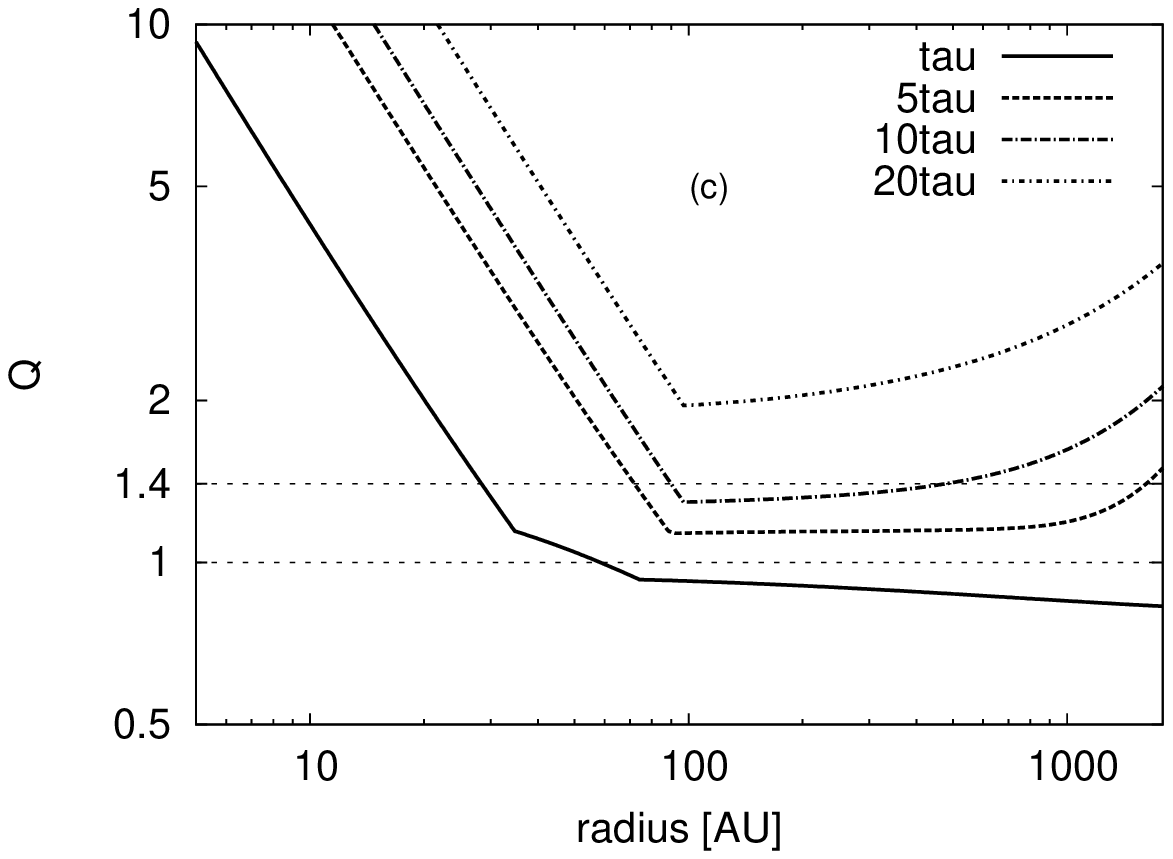}
\FigureFile(80mm,80mm){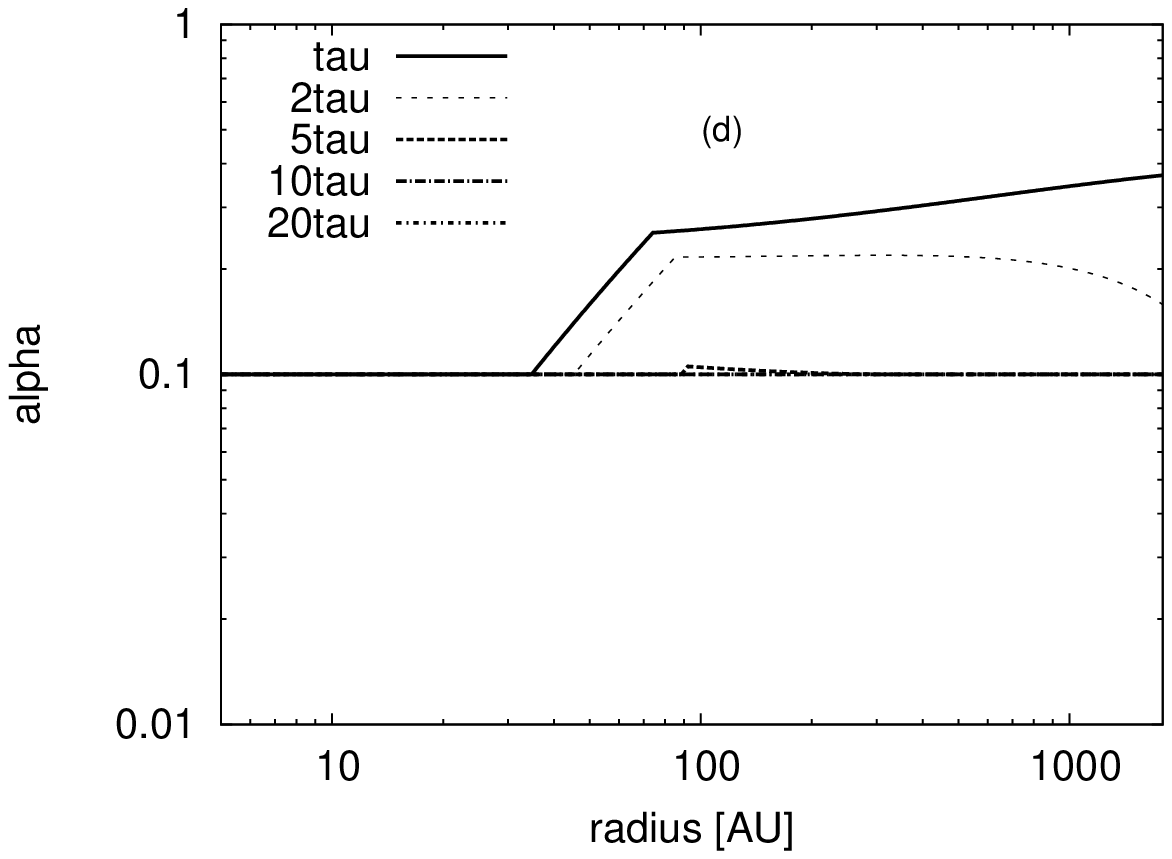}

\caption{Radial profiles of the surface density $\Sigma$ (a), mass accretion rate (b), Toomre's Q-value (c), and effective viscous parameter $\alpha$ (d), for the case with $\alpha_0 =0.1$ and $\omega=0.3$. The mass accretion rate onto the disk after $\tau$ is set to be zero to represent the depletion of the parent cloud.}
\end{center}
\end{figure}
In Figure 6, time evolutions of the disk with the effect of decline of mass infall rate onto the disk are shown. 
Depletion effect of the parent cloud core is modeled as equation (\ref{depletion}). 
For temperature distribution, barotropic equation of state (equation (\ref{soundspeedbaro})) is assumed. 
Distributions of the surface density and mass accretion rate in several time epochs are shown in Figure 6 (a) and (b), respectively. 
As seen in Figure 6 (a) and 6 (b), the surface density near the center and mass accretion rate onto the star decreases after the depletion time $\tau=63000~\rm yr$. 
In Figure 6 (a), area with $\Sigma\propto r^{-0.75}$ appears from the inner radius as in Figure 2. 
This result indicates that power index of radial distribution of the surface density near the center does not depend on the time dependence of the mass flux onto the disk. 
From Figure 6 (c), it is seen that after 5$\tau$ $Q$-value is larger than unity and the disk is gravitationally stable. 
In Figure 6 (d), the effective $\alpha$ is found to become spatially constant after 5$\tau$ owing to $\alpha_{\rm G}\sim 0$ in equation (\ref{alpha}). 
Disk is stabilized in $t> 5\tau$ because sufficient amount of mass in the disk is accreted by the central star. 
\begin{figure}
\begin{center}
\FigureFile(80mm,80mm){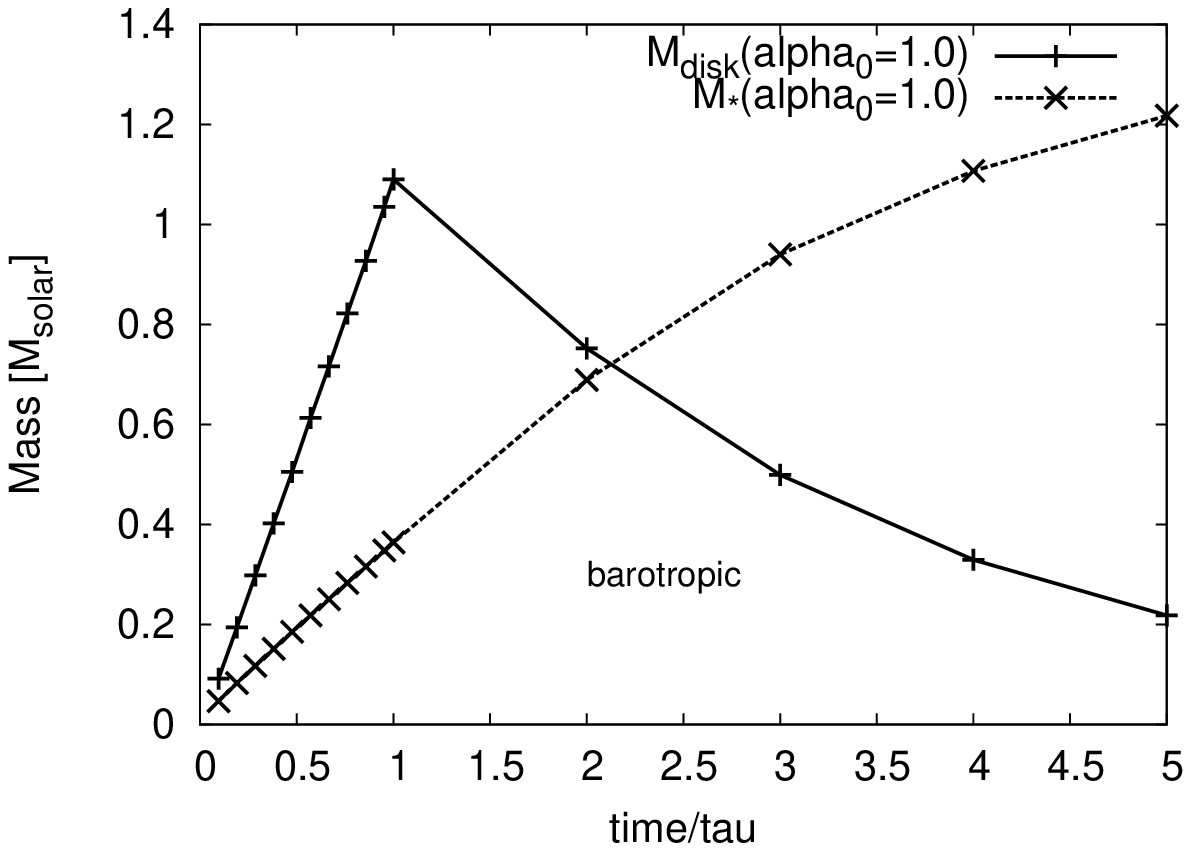}
\FigureFile(80mm,80mm){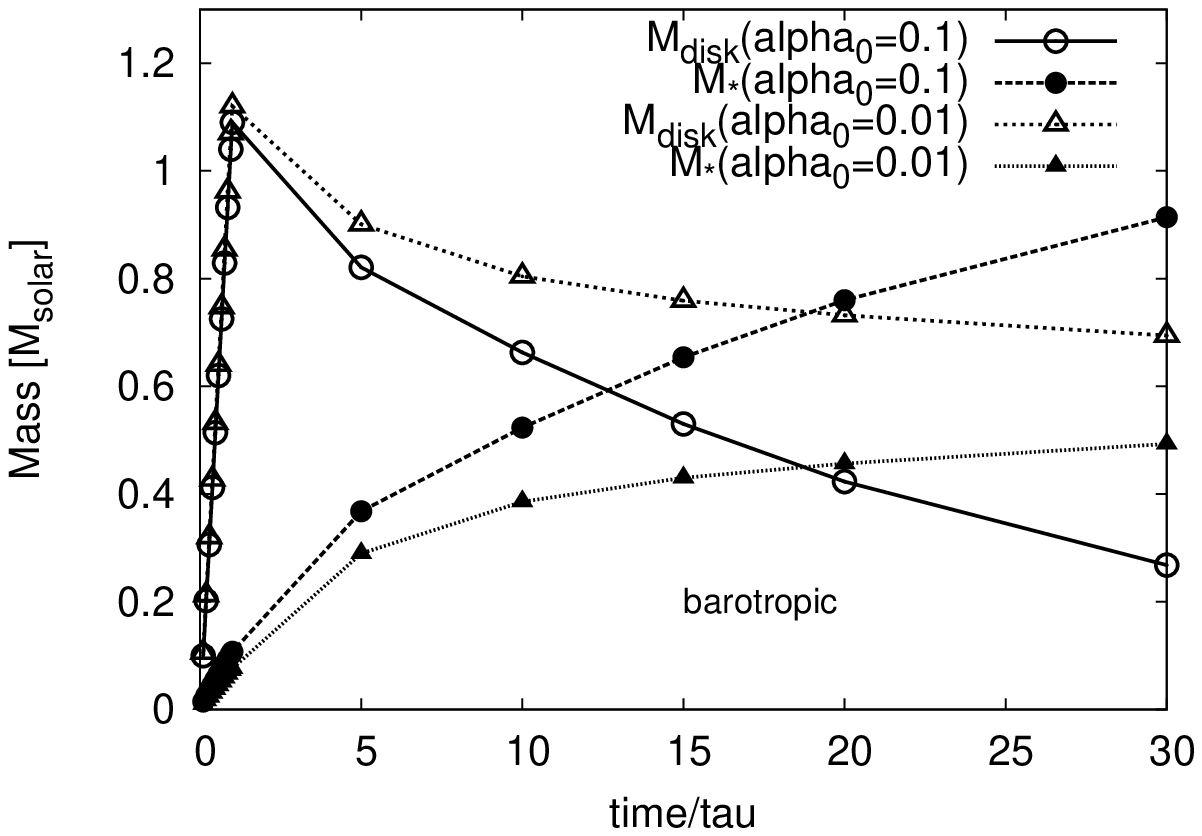}
\FigureFile(80mm,80mm){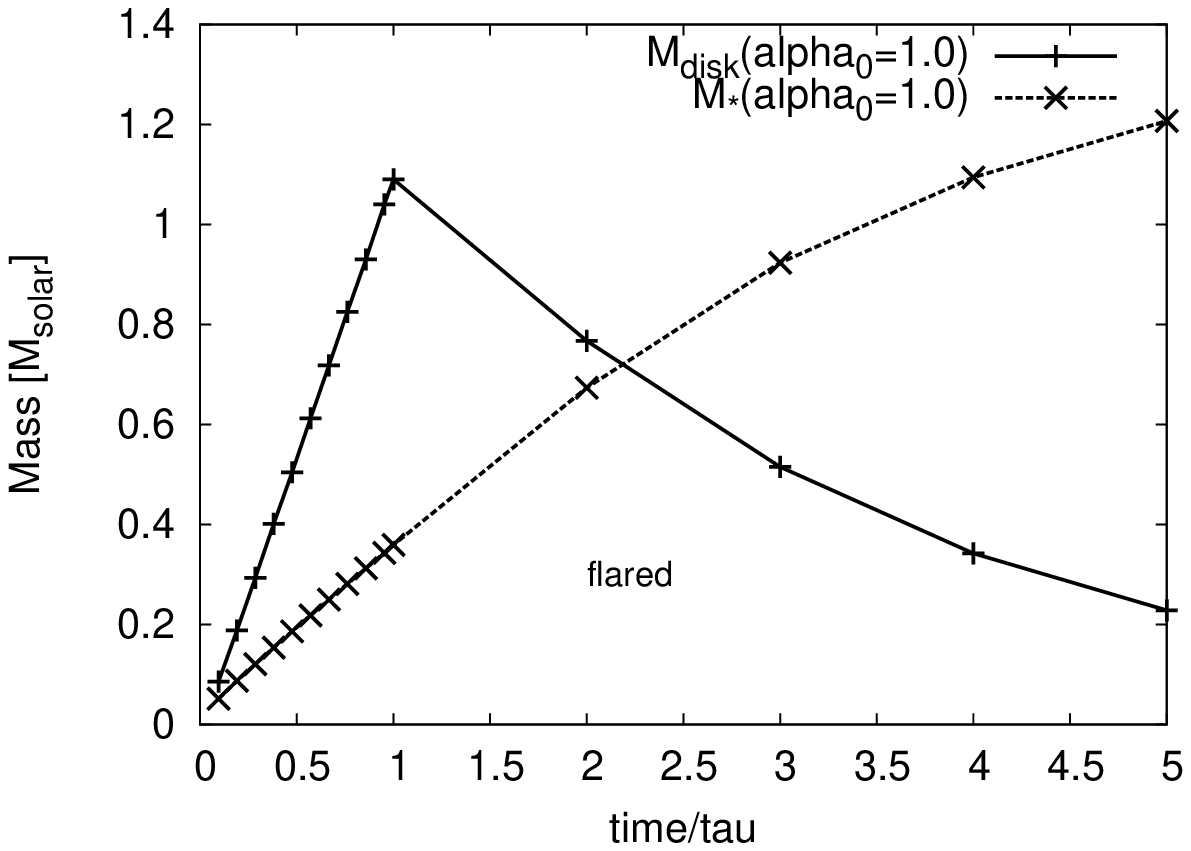}
\FigureFile(80mm,80mm){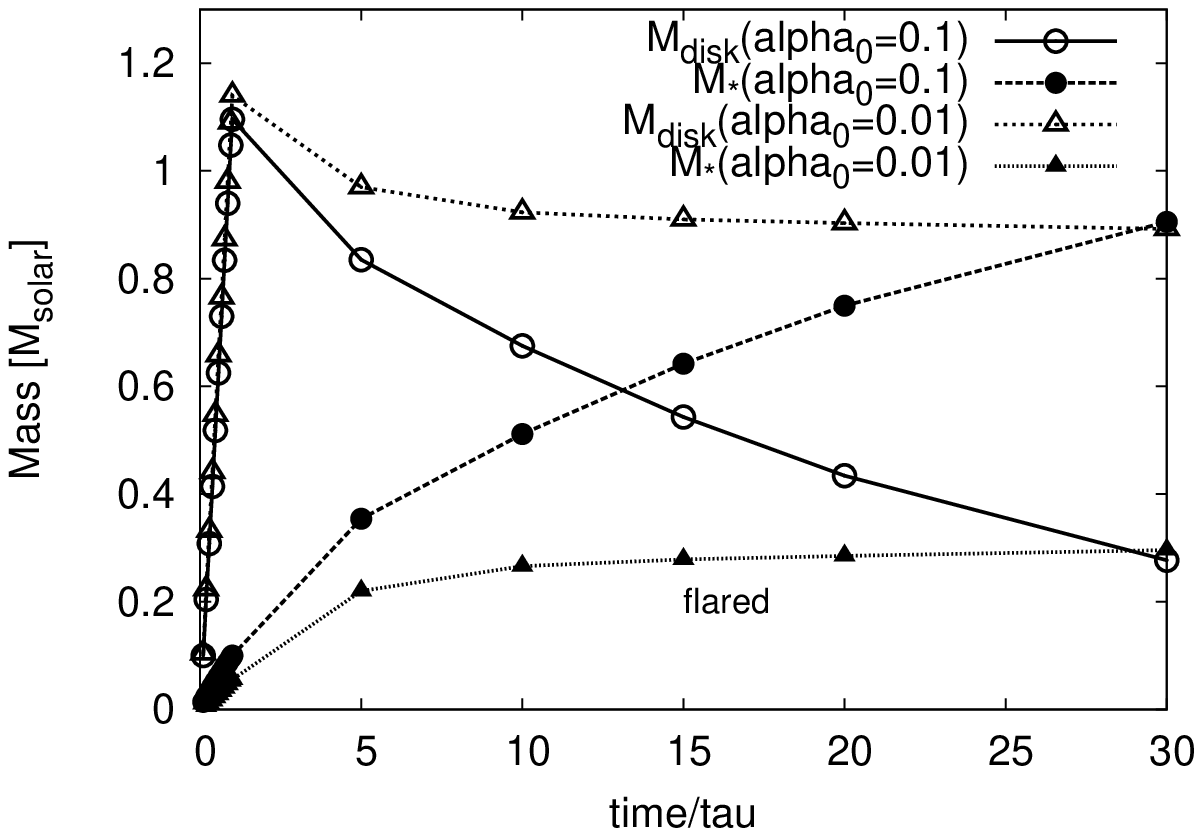}
\caption{The time evolutions of the disk mass and the star mass for the case with barotropic equation of state (left top and right top). The time evolutions of the disk mass and the star mass for the case with the flared disk (left bottom and right bottom).}
\end{center}
\end{figure}

In Figure 7, time evolution of $M_{\rm disk}$ and $M_*$ is shown.
It is seen that mass of the disk tends to be larger than mass of the star in $t<\tau$ and that 
the disk-to-star mass ratio $M_{\rm disk}/M_*$ decreases in $t>\tau$. 
It is also seen that $M_{\rm disk}/M_*$ is smaller than unity in $t> 2$--$3\tau \simeq 10^5~\rm yr$ in the case with $\alpha_0 =1$ and in $t>15$--$20\tau \simeq 10^6~\rm yr$ in the case with $\alpha_0 =0.1$. 
However, in the case with small $\alpha_0=0.01$, it is seen that $M_{\rm disk}/M_*$ is still larger than unity at $t=2~\rm Myr~(\sim 40\tau)$. 

In summary, in this subsection it is found that the disk-to-star mass ratio begins to decrease by the effect of decline of the mass infall rate onto the disk. 



\section{Discussion}

\subsection{Possible Surface Density Profile as a Result of Viscous Evolution}
As the result of viscous accretion, mass accretion rate $\dot{M}$ approaches spatially constant from the inner region. 
We call this state the quasi-steady state. 
As shown in \S 3, 
the radial profile of the surface density $\Sigma$ in the flared disk model approaches the different profile from the other temperature model. 
Here, we analytically derive the radial profile of the surface density in the quasi-steady state near the center. 
In this subsection, the viscous parameter $\alpha$ is assumed to be spatially constant. 

According to equation (\ref{diffusiveeq}), two types of steady states are derived by the conditions 
\begin{equation}
\nu \Sigma r^3\frac{\partial \Omega}{\partial r}={\rm constant},
\label{stable1}
\end{equation}
or
\begin{equation}
\frac{1}{\frac{\partial(rv_{\phi})}{\partial r}}\frac{\partial}{\partial r}(\nu \Sigma r^3\frac{\partial \Omega}{\partial r})={\rm constant}.
\label{stable2}
\end{equation}
Equation (\ref{stable1}) means that the torque is spatially constant, 
and mass accretion $\dot{M}=0$. 
Equation (\ref{stable2}) means that mass accretion rate is spatially constant. 

Assume that the mass within $r$ is described as $M(r)=M_1 r^n$. 
According to the assumed form of $M(r)$, 
the case with $n=0$ corresponds to the case where the mass of the disk is negligible. 
On the other hand, the case with $n=1$ corresponds to the disk without the central star. 
The condition of the surface density $\Sigma$ for the steady state is found to be $\Sigma\propto r^{-2}$ from (\ref{stable1}), and $\Sigma \propto r^{\frac{n-3}{2}}$ from (\ref{stable2}) in the isothermal case. 
Using equation (\ref{stable2}), $\Sigma \propto r^{-1.5}$ is derived for $n=0$ and $\Sigma\propto r^{-1}$ is derived for $n=1$. 
As shown in \S 2.4, it is found that the power index of the radial profile of the surface density is $r^{-1.5}$ for $r\rightarrow 0$. 
This indicate that $r^{-1.5}$ in the quasi-steady state is given by equation (\ref{stable2}) instead of (\ref{stable1}). 

In the non-isothermal case with barotropic equation of state (\ref{soundspeedbaro}), 
the radial dependence of sound speed is given by 
\begin{equation}
c_{\rm s} \propto (\Sigma\Omega)^{1/6}
\label{cstenkai}
\end{equation} 
in the case with $\gamma=7/5$. 
This form of sound speed affects on the time evolution of $\Sigma$ via $\nu$. 
According to the assumed form of $M(r)$, equation (\ref{cstenkai}) is written as 
\begin{equation}
c_{\rm s}\propto r^{-3/8}(\propto T^{-3/4}) 
\label{barostea}
\end{equation}
in the case with $n=0$. 
This profile of sound speed is the same as the profile in the flat disk model. 
This is why the power index on $\Sigma$ is the same between Figure 2 (a) and Figure 3 (b). 
Using equations (\ref{stable2}) and (\ref{barostea}), $\Sigma \propto r^{-3/4}$ is derived with $n=0$ and $\gamma=7/5$. 
This profile of the surface density is seen in the inner region in \S 3.1.1. 

KI02 indicates that the radial profile of mass of protoplanets qualitatively changes when the power index of the surface density of planetesimal is smaller than $-2$. 
Condition (\ref{stable2}) indicates the profile 
\begin{equation}
\Sigma \propto r^{\frac{3(\gamma-3)}{2(3\gamma-1)}}(\equiv r^A)
\label{stablebaro}
\end{equation}
for the case with $n=0$. 
In the range $1<\gamma<7/5$ treated in this paper, equation (\ref{stablebaro}) indicates that the power index of $\Sigma$ distribution becomes $-3/2<A<-3/4$. 
All the results in our numerical calculations are also within in this range. 
Thus, it is indicated that power index $|A|$ is expected to be smaller than $\beta=2$ in $\Sigma \propto r^{-\beta}$ in our model. 

For reference, in the MMSN model of protoplanetary disk (\cite{HayashiPPD}), 
radial profile of sound speed is $c_{\rm s} \propto r^{-1/4}$. 
In the corresponding flared disk model in this paper, 
in the steady state for $n=0$, the equation (\ref{stable1}) gives $\Sigma \propto r^{-1.5}$ as in the MMSN disk. 
On the other hand, from the equation (\ref{stable2}) different profile $\Sigma \propto r^{-1}$ is derived. 
This profile of the surface density is seen in the inner region of Figure 3 (a). 
\subsection{Comparison with the Previous Studies}
Different from the results in Nakamoto \& Nakagawa (\yearcite{NN1995}), 
in our model the disk-to-star mass ratio $M_{\rm disk}/M_*$ tend to be larger than unity even with the large $\alpha(\sim 1)$, 
as long as the temporary constant mass flux $8.512\frac{c_{\rm s}^3}{G}$ onto the disk is assumed. 
This is because in this paper higher mass accretion rate onto the disk is assumed base on SH98 
solution which describes dynamical collapse. 

HG05 used one-dimensional theoretical model of formation and viscous evolution 
of the circumstellar disk in order to compare with observed disks around T-Tauri stars. 
They used temporary constant $\alpha$ in the viscous evolution of the disk from class 0 to T-Tauri phase. 
Their model requires $\alpha\sim 10^{-2}$ in order to explain the luminosity of the T-Tauri phase. 
This value is too small to provide mass accretion rate effectively provided by gravitational instability in the early phase of disk formation. 
Furthermore, they assumed that the cloud envelope is spherically symmetric and they used 
the solution by Shu (\yearcite{shu1977}) as the temporary constant mass flux onto the disk. 
In their result with $\alpha\sim 10^{-2}$, the disk-to-star mass ratio tend to be smaller than unity in several Myr 
after the beginning of the accretion phase. 
In the view point of the disk-to-star mass ratio, their results are similar to Nakamoto \& Nakagawa (\yearcite{NN1995}) and different from ours. 
This is because mass accretion rate onto the disk in HG05 is smaller than that in our model. 

Mass accretion rate onto the disk is expected to depend on whether 
a molecular cloud core collapses dynamically or not.
Although actual value of accretion rate onto the disk will depend on 
the complex structure of cloud and on turbulent velocity fields, 
the assumption of dynamical collapse adopted in this paper will tend to be applicable 
in cases where a cloud start to collapse from a magnetically super-critical mass. 
Indeed, in three-dimensional calculations presented by Inutsuka et al. (\yearcite{Inuetal2010}) and Machida et al (\yearcite{Machietal2010}), 
it is reported that the disk mass is larger than the star mass in $t<10^4 ~\rm yr$ from the protostar formation in the case with large rotational energy in the molecular cloud. 
In the two-dimensional calculation by Vorobyov (\yearcite{Voro2010b}), it is also reported that the disk mass become comparable to the star mass in $t<10^5~\rm yr$ after the beginning of the accretion phase. 
Results in the present paper with $\alpha_0~=~0.1$--$1$ are consistent with these numerical studies. 
  
\section{Summary}
In this paper, 
we studied the unsteady viscous disk evolution subject to mass loading from dynamically collapsing envelope. 
Simultaneous evolution of a small protostar and surrounding disk is investigated. 
In the present model, mass accretion is treated simply by effective viscosity $\alpha$. 
The gravitational force from the disk itself is taken into account approximately as equation (\ref{gravi}) 
and its effect is taken into disk angular speed $\Omega$. 
Three types of temperature distribution by equations (\ref{soundspeedbaro}, \ref{soundspeedirrad}) are considered. 
Our results are summarized as follows: 

1. The profile of the surface density in the inner region is found to approach the power-law distribution with power-index determined mainly by the process of angular momentum transport. 
The difference between three cases with different temperature distributions is introduced through viscosity $\nu =\alpha (c_{\rm s})^2/\Omega$. 
In the quasi-steady state, profile of the surface density is given by equation (\ref{stablebaro})
and $\Sigma \propto r^{\alpha}~{\rm with}~-2<\alpha<-1$ is typically indicated. 
The radial profiles of the surface density, azimuthal velocity, and mass accretion rate are divided into three branches. 
These branches are divided by two characteristic radii, $r_{\rm visc}$ outside that the disk is gravitationally unstable and $r_{\rm temp}$ inside that the disk is non-isothermal. 

2. It is found that the disk-to-star mass ratio $M_{\rm disk}/M_*$ tends to be larger than unity in almost all cases with $\alpha<1$, as long as steady dynamical flow onto the disk given by SH98 is assumed. 
While it is found that the disk-to-star mass ratio decreases with time after the time epoch when 
accretion rate onto the disk declines with time due to the depletion of the parent cloud core. 
Numerical results show that 
$M_{\rm disk}/M_*$ is smaller than unity after $10^5~\rm yr$ from the beginning of the accretion phase in the case with $\alpha_0 =1$ and after $10^6~\rm yr$ in the case with $\alpha_0 =0.1$. 
$M_{\rm disk}/M_*$ is still larger than unity in the case with $\alpha_0=0.01$ at $t=2~\rm Myr $.

\bigskip
We thank an anonymous referee for useful comments and suggestions that helped improve this paper. 
We are grateful to Fumio Takahara for fruitful discussion and continuous encouragement. 
We also acknowledge useful discussions with Shigeo S. Kimura.




\appendix


\section{The value of $\Sigma_1$ in our model}
 In our model, surface density profile and angular momentum distribution in SH98 is used as initial condition. 
In the solution for accretion phase in Saigo \& Hanawa (\yearcite{SH98}), the radius of the shock front is written as $Ac_{\rm{s}}t$. 
Mass accretion rate $\dot{M}$ is written as
\begin{equation}
\frac{M_{\rm d}}{t}=\mu_{\rm d} \frac{c_{\rm s}^3}{G},
\label{Md1}
\end{equation}
where $M_{\rm d}$ corresponds $M_{\rm disk}$ in our model and $\mu_{\rm d}$ is $m_0$ in our model. 
These two coefficients $A$ and $\mu_{\rm d}$ depend on the parameter of rotation $\omega$. 
Values of $\mu_{\rm d}$ and $A$ in SH98 are shown in Table 1. 
Using equation (\ref{gravi}), $M_{\rm disk}$ is also written as 
\begin{equation}
\int_{0}^{r_{\rm sh}}2\pi r\Sigma_1 r^{-1}dr=2\pi\Sigma_1 r_{\rm s}=2\pi\Sigma_1Ac_{\rm s}t. 
\label{Md2}
\end{equation}
From equations (\ref{Md1}) and (\ref{Md2}), $\Sigma_1$ is written as
\begin{equation}
\Sigma_1 =\frac{c_{\rm s}^2}{2\pi G}\frac{\mu_{\rm d}}{A}.
\end{equation}
\begin{table}
  \caption{The value of coefficient $\mu_{\rm d}$ and $A$. (c.f., Saigo \& Hanawa (1998))}\label{tab: table2}
  \begin{center}
    \begin{tabular}{ccc}
      \hline
     $\omega$ & $\mu_{\rm d}$ & $A$ \\
      \hline
     0.2&9.759&0.363\\
     0.3&8.512&0.676\\
     0.4&6.682&0.986
    \end{tabular}
  \end{center}
\end{table}

\section{Validity of the Present Model}
First, as mentioned in \S 2.2, 
we assumed that the pressure gradient force is negligible. 
Thus, it is assumed that the centrifugal force only balances with the gravitational force in our model. 
We estimate the radius where the pressure gradient force is not negligible in the non-isothermal disk with barotropic equation of state. 
Using the initial condition of the surface density distribution $\Sigma=\Sigma_1 r^{-1}$, 
it is found that the pressure gradient force is larger than the gravitational force inside $0.962~{\rm AU}$. 
Thus, the pressure gradient force is smaller than the gravitational force outside $5~{\rm AU}$, 
which is our inner boundary radius $r_{\rm ib}$. 

Second, as mentioned in \S 2.2, we used $|v_{\rm r}|<<v_{\phi}$. 
This condition corresponds to the fact radial velocity induced by viscous flow is sufficiently slow. 
This condition is achieved when $t_{\rm diff}(r)>t_{\rm rot}(r)$, 
where $t_{\rm diff}=\frac{r^2}{\nu}$ is the diffusion time and $t_{\rm rot}=1/\Omega$ is the rotation time, respectively. 
We estimate the radius $r_{\rm rot}$ where $t_{\rm diff}$ equals to $t_{\rm rot}$. 
Using the initial condition of the surface density distribution $\Sigma_1 r^{-1}$, 
$t_{\rm rot}$ is written as $6.67\times (r/1{\rm AU})~{\rm yr}$. 
Then the condition $t_{\rm diff}(r)>t_{\rm rot}(r)$ is written as $r^{2/3}>\alpha/0.947$. 
Thus, $r_{\rm rot}\simeq \alpha^{3/2}~{\rm AU}$. 
This $r_{\rm rot}$ is safely smaller than the inner boundary radius $r_{\rm ib}$ in this paper with $\alpha<1$. 

Third, as mentioned in \S 2.3, 
we used the shock front radius $r_{\rm sh}$ from SH98 without viscosity. 
To use $r_{\rm sh}$ safely, $t_{\rm SH}(r)$ must be smaller than $t_{\rm diff}$, where the $t_{\rm SH}$ is the time when the shock front given by SH98 reaches radius $r$. 
In the region with $t_{\rm diff}(r)<t_{\rm SH}(r)$, 
mass that falls onto the disk is immediately accreted by the central star. 
We estimate the radius $r_{\rm eq}$ where $t_{\rm diff}$ equals to $t_{\rm SH}$. 
The radius $r_{\rm eq}$ is written as $13.11\alpha^{3/2}~{\rm AU}$ for the case with $\omega=0.3$, and $\rho_{\rm crit}=5.0\times10^{-14}~{\rm g}/{\rm cm}^2$. 
The disk inside of the radius $r_{\rm eq}$ will have the different initial state from our model (equation (\ref{initialSigma})). 
In this sense our initial surface density profile by equation (\ref{initialSigma}) is overestimated inside $r_{\rm eq}$. 

We adopted the thin-disk approximation in this paper. 
However, this approximation is invalid for systems with $M_{\rm disk}/M_{\rm star}>1$. 
This is because the scale height $h$ become greater than the radial distance $r$ 
(e.g., Vorobyov \yearcite{Voro2009}). 
The aspect ratio $A(r)=\frac{h}{r}$ for Keplerian disk is usually approximated by
\begin{equation}
A(r)=CQM_{\rm disk}/M_*,
\end{equation}
where $C\sim 1$ is constant. 
$A(r)$ is larger than $0.1$ if disk-to-star mass ratio is larger than unity. 
This effect is not taken into account in the present paper. 



\begin{thebibliography}{}
\bibitem[Balbus \& Hawley (1991)]{BH1991}
Balbus, S. A., Hawley, J. F. 1991, ApJ, 376, 214
\bibitem[Balbus \& Hawley(1998)]{BalbusHawley}
Balbus, S. A., \& Hawley, J. F. 1998, Rev. Mod. Phys., 70, 1 
\bibitem[Beckwith et al.(1990)]{Bec}
Beckwith, S. V. W., Sargent, A. I., Chini, R. S., \& G\"{u}sten, R. 1990, Astron. J., 99, 924
\bibitem[Boley et al. (2006)]{Boley}
Boley, A. C., Mej\'{i}a, A. C., Durisen, R. H., Cai,K., Pickett, M. K., \& D'Alessio, P. 2006, ApJ, 651, 517
\bibitem[Bontemps et al. (1996)]{Bontemps}
Bontemps, S. ,Andr\'{e}, P, Terebey, S., \& Cabrit, S. 1996, A\&A, 311,858
\bibitem[Hartmann (2008)]{Hartmann}
Hartmann, L. 2008, Accretion Processes in Star Formation, Second Edition,Cambridge Astrophysics Series, 47
\bibitem[Hayashi, Narita, \& Miyama (1982)]{Hayashidisk}
Hayashi, C., Narita, S., \& Miyama, S. M. 1982, Prog. Theor. Phys., 68, 1949
\bibitem[Hayashi et al.(1985)]{HayashiPPD}
Hayashi, C., Nakazawa, and Y.Nakagawa 1985. Formation of the solar System. In Protostars and Planets II, pp1100-1153. Univ. of Arizona Press, Tucson.
\bibitem[Hueso \& Guillot (2005)]{HuesoGuillot}
Hueso, G., \& Guillot, T. 2005, A \& A, 442, 703 
\bibitem[Hunter(1977)]{Hunter}
Hunter, C. 1977, ApJ, 218, 834
\bibitem[Inutsuka et al. (2010)]{Inuetal2010}
Inutsuka, S., Machida, M. N., \& Matsumoto, T 2010, ApJL, 718, L58
\bibitem[Kokubo \& Ida (2002)]{KI2002}
Kokubo, E., Ida, S. 2002, ApJ, 581, 666
\bibitem[Kratter et al.(2010)]{Kratter}
Kratter, K. M., Matzner, C. D., Krumholz, M. R., \& Klein, R. I. 2010, ApJ, 708, 1585
\bibitem[Larson(1969)]{Larson}
Larson, R. 1969, MNRAS, 145, 271
\bibitem[Lee (2010)]{Lee}
Lee, C. F. 2010. ApJ, 725, 712
\bibitem[Lin \& Pringle (1987)]{LP87}
Lin, D. N. C., \& Pringle, J. E. 1987, MNRAS, 225, 607
\bibitem[Lin \& Pringle (1990)]{LP90}
Lin, D. N. C., \& Pringle, J. E. 1990, ApJ, 358, 515
\bibitem[Machida et al. (2010)]{Machietal2010}
Machida, M. N, Inutsuka, S., \& Matsumoto, T. 2010, ApJ, 724, 1006
\bibitem[Masunaga \& Inutsuka(2000)]{MasunagaInutsuka2000}
Masunaga, H., \& Inutsuka, S. 2000, ApJ, 531, 350
\bibitem[Masunaga, Miyama \& Inutsuka(1998)]{MasuInu1998}
Masunaga, H., Miyama, S. M., \& Inutsuka, S. 1998, ApJ, 495, 346
\bibitem[Matsumoto et al.(1997)]{Matsu}
Matsumoto, T., Hanawa, T., \& Nakamura, F. 1997, ApJ, 478,569
\bibitem[Mestel(1963)]{Mestel}
Mestel, L. 1963, MNRAS, 126, 553
\bibitem[Nakamoto \& Nakagawa (1994,~1995)]{NaNa1994}
Nakamoto, T., \& Nakagawa, Y. 1994, ApJ, 421, 640
\bibitem[Nakamoto \& Nakagawa (1995)]{NN1995}
Nakamoto, T., \& Nakagawa, Y. 1995, ApJ, 445, 330
\bibitem[Narita, Hayashi \& Miyama (1984)]{NHM84}
Narita, S., Hayashi, C., \& Miyama, S. M. 1984, Prog. Theor. Phys., 72, 1118
\bibitem[Nomura \& Mineshige(2000)]{NoMine}
Nomura, H., \& Mineshige, S. 2000, ApJ, 536, 429
\bibitem[Penston(1969)]{Penston}
Penston, M., V. 1969, MNRAS, 144, 425
\bibitem[Pringle (1981)]{Pringle}
Pringle, J. E., Ann. Rev. Astron. Astrophys. 1981, 19, 137
\bibitem[Rice et al. (2010)]{Rice2010}
Rice, W. K. M., Mayo, J. H., \& Armitage, Philip J. 2010, MNRAS, 402, 1740
\bibitem[Saigo \& Hanawa(1998)]{SH98}
Saigo, K., \& Hanawa, T. 1998, ApJ, 493, 342
\bibitem[Shakura \& Sunyaev(1971)]{Shakura}
Shakura, N. I., \& Sunyaev, R. A. 1973, A\&A, 24, 337
\bibitem[Shu(1977)]{shu1977}
Shu, F. H. 1977, ApJ, 214, 488
\bibitem[Shu,Adams \& Lizano(1987)]{shu1987}
Shu, F. H., Adams, F. C., \& Lizano, S. 1987, Ann. Rev. Astron. Astrophysics, 25, 23-81
\bibitem[Toomre(1964)]{ToomreQ}
Toomre, A. 1964, ApJ, 139, 1217
\bibitem[Tsuribe(1999)]{Tsu}
Tsuribe, T. 1999, ApJ, 527, 102
\bibitem[Visser et al. (2009)]{Visser}
Visser, R., van Dishoeck, E. F., Doty, S. D., \& Dullemond, C. P. 2009, A\&A, 495, 881
\bibitem[Vorobyov (2009)]{Voro2009}
Vorobyov, E. I. 2009, New Astron., 15, 24
\bibitem[Vorobyov (2010a)]{Voro2010a}
Vorobyov, E. I. 2010a, ApJ, 713, 1059
\bibitem[Vorobyov (2010b)]{Voro2010b}
Vorobyov, E. I. 2010b, ApJ, 723, 1294
\bibitem[Vorobyov \& Basu(2007,2009)]{VoroBasu2007}
Vorobyov, E. I., \& Basu, S. 2007, MNRAS, 381, 1009
\bibitem[Voroby \& Ba(2009)]{VB2009}
Vorobyov, E. I., \& Basu, S. 2009, MNRAS, 393, 822

\end{thebibliography}
\end{document}